\documentstyle[12pt,graphicx,subfigure]{article}
\begin{document}

\def\AEF{A.E. Faraggi}
\def\NPB#1#2#3{{\it Nucl.\ Phys.}\/ {\bf B#1} (#2) #3}
\def\NPA#1#2#3{{\it Nucl.\ Phys.}\/ {\bf A#1} (#2) #3}
\def\PLB#1#2#3{{\it Phys.\ Lett.}\/ {\bf B#1} (#2) #3}
\def\PRD#1#2#3{{\it Phys.\ Rev.}\/ {\bf D#1} (#2) #3}
\def\PRL#1#2#3{{\it Phys.\ Rev.\ Lett.}\/ {\bf #1} (#2) #3}
\def\PRT#1#2#3{{\it Phys.\ Rep.}\/ {\bf#1} (#2) #3}
\def\MODA#1#2#3{{\it Mod.\ Phys.\ Lett.}\/ {\bf A#1} (#2) #3}
\def\IJMP#1#2#3{{\it Int.\ J.\ Mod.\ Phys.}\/ {\bf A#1} (#2) #3}
\def\nuvc#1#2#3{{\it Nuovo Cimento}\/ {\bf #1A} (#2) #3}
\def\RPP#1#2#3{{\it Rept.\ Prog.\ Phys.}\/ {\bf #1} (#2) #3}
\def\APJ#1#2#3{{\it Astrophys.\ J.}\/ {\bf #1} (#2) #3}
\def\APP#1#2#3{{\it Astropart.\ Phys.}\/ {\bf #1} (#2) #3}
\def\etal{{\it et al\/}}

\newcommand{\bev}{\begin{verbatim}}
\newcommand{\beq}{\begin{equation}}
\newcommand{\beqa}{\begin{eqnarray}}
\newcommand{\beqn}{\begin{eqnarray}}
\newcommand{\eeqn}{\end{eqnarray}}
\newcommand{\eeqa}{\end{eqnarray}}
\newcommand{\eeq}{\end{equation}}
\newcommand{\Eev}{\end{verbatim}}\newcommand{\bec}{\begin{center}}
\newcommand{\eec}{\end{center}}
\def\ie{{\it i.e.}}
\def\eg{{\it e.g.}}
\def\half{{\textstyle{1\over 2}}}
\def\nicefrac#1#2{\hbox{${#1\over #2}$}}
\def\third{{\textstyle {1\over3}}}
\def\quarter{{\textstyle {1\over4}}}
\def\m{{\tt -}}
\def\mass{M_{l^+ l^-}}
\def\p{{\tt +}}

\def\slash#1{#1\hskip-6pt/\hskip6pt}
\def\slk{\slash{k}}
\def\GeV{\;{\rm GeV}}
\def\TeV{\;{\rm TeV}}
\def\y{\;{\rm y}}

\def\l{\langle}
\def\r{\rangle}
\newcommand{\lsim}   {\mathrel{\mathop{\kern 0pt \rlap
  {\raise.2ex\hbox{$<$}}}
  \lower.9ex\hbox{\kern-.190em $\sim$}}}
\newcommand{\gsim}   {\mathrel{\mathop{\kern 0pt \rlap
  {\raise.2ex\hbox{$>$}}}
  \lower.9ex\hbox{\kern-.190em $\sim$}}}
\renewcommand{\thefootnote}{\fnsymbol{footnote}}
\setcounter{footnote}{0}

\begin{titlepage}
\samepage{
\setcounter{page}{1}

\rightline{OUTP-03-22P}
\rightline{\tt hep-ph/0308169}
\rightline{August 2003}
\vspace{0.5cm}
\begin{center}
 {\Large \bf
Large Scale Air Shower Simulations\\
 and the Search for New Physics at AUGER\\}
\vspace{1.5cm}
 {\large
Alessandro Cafarella$^1$\footnote{Alessandro.Cafarella@le.infn.it},
Claudio Corian\`{o}$^1$\footnote{Claudio.Coriano@le.infn.it} and
Alon E. Faraggi$^{2}$\footnote{faraggi@thphys.ox.ac.uk}}

\vspace{.25cm}
{\it $^1$Dipartimento di Fisica,
 Universita' di Lecce,\\
 I.N.F.N. Sezione di Lecce,
Via Arnesano, 73100 Lecce, Italy\\}
\vspace{.25cm}
{\it $^2$Theoretical Physics Department,\\
University of Oxford, Oxford, OX1 3NP, United Kingdom\\}
\vspace{.25cm}

\end{center}
\begin{abstract}
Large scale airshower simulations around the GZK cutoff are
performed. An extensive analysis of the behaviour
of the various subcomponents of the cascade is presented.
We focus our investigation both
on the study of total and partial multiplicities along the entire
atmosphere and on the geometrical structure of the various
cascades, in particular on the lateral distributions.
The possibility of detecting new physics in Ultra High Energy
Cosmic Rays (UHECR) at AUGER is
also investigated. We try to disentangle
effects due to standard statistical fluctuations in the first proton
impact in the shower formation from the underlying interaction and comment on
these points.
We argue that theoretical models predicting large missing energy may have a
chance
to be identified, once the calibration errors in the energy measurements
are resolved by the experimental collaborations, in measurements
of inclusive multiplicities.

\end{abstract}
\smallskip}
\end{titlepage}

\section{Introduction}

One of the most intriguing experimental observations of recent years
is the detection of Ultra--High--Energy--Cosmic--Rays (UHECR),
with energy in excess of the Greisen--Zatsepin--Kuzmin (GZK)
cutoff \cite{uhecr} (for a review see \cite{la}).
While its validity is still under some dispute,
it is anticipated that the forthcoming AUGER \cite{auger}
and EUSO \cite{euso} experiments will provide enough statistics to 
resolve the debate. From a theoretical perspective, the Standard Model of
particle physics and its Grand Unified extensions indicate
that many physical structures may lie far beyond the reach
of terrestrial collider experiments. If this eventuality
materializes it may well be that the only means of unlocking the
secrets of the observed world will be mathematical rigor and peeks
into the cosmos in its most extreme conditions. In this
context the observation of UHECR is especially puzzling
because of the difficulty in explaining the events without
invoking some new physics.
There are apparently no astrophysical sources
in the local neighborhood that can account for the
events. The shower profile of the
highest energy events is consistent with identification of the
primary particle as a hadron but not as a photon or a neutrino.
The ultrahigh energy events observed in the air shower arrays
have muonic composition indicative of hadrons.
The problem, however, is that the propagation of hadrons
over astrophysical distances is affected by the
existence of the cosmic background radiation, resulting
in the GZK cutoff on the maximum energy of cosmic ray
nucleons $E_{\rm GZK}\le10^{20}\;{\rm eV}$ \cite{gzk}.
Similarly, photons of such high energies have a mean free path of less
than 10 Mpc due to scattering {}from the cosmic background radiation and
radio photons. Thus, unless the primary is a neutrino,
the sources must be nearby. On the other hand, the primary
cannot be a neutrino because the neutrino interacts very weakly
in the atmosphere. A neutrino primary would imply that the
depths of first scattering would be uniformly distributed
in column density, which is contrary to the observations.

The most exciting aspect of the UHECR
is the fact that the AUGER and EUSO
experiments will explore the physics associated with these events,
and provide a wealth of observational data.
Clearly, the first task of these experiments is to establish
whether the GZK cutoff is violated, and to settle the
controversy in regard to the air shower
measurement.

\section{Probing new physics with UHECR}
We may, however, entertain the
possibility that these experiments can probe
various physics scenarios. In the first place,
the center of mass energy in the collision
of the primary with the atmosphere is of the order
of 100 TeV and exceed the contemporary, and forthcoming,
collider reach by two orders of magnitude.
Thus, in principle the air shower analysis should
be sensitive to any new physics that is assumed
to exist between the electroweak scale and the
collision scale due to the interaction of the
primaries with the atmospheric nuclei. Other
exciting possibilities include the various
explanations that have been put forward to
explain the existence of UHECR events \cite{explanations,berez, ben, sb, cfp},
and typically assume some form of new physics.
One of the most intriguing possible solutions
is that the UHECR primaries originate {}from the decay of long--lived
super--heavy relics, with mass of the order of $10^{12-15}\;{\rm GeV}$
\cite{berez, ben, sb, cfp}.
In this case the primaries for the observed UHECR would originate
from decays in our galactic halo, and the GZK bound would not apply.
This scenario is particularly interesting due to the possible
connection with superstring theory.
{}From the particle physics perspective the meta--stable super--heavy
candidates should possess several properties.
First, there should exist a stabilization mechanism which produces
the super--heavy state with a lifetime of the order of
$
10^{17}s\le \tau_X \le 10^{28}s,~
$
and still allows it to decay and account for the observed UHECR events.
Second, the required mass scale of the meta--stable state
should be of order, $M_X~\sim~10^{12-13}{\rm GeV}.$
Finally, the abundance of the super--heavy relic
should satisfy the relation
$
({\Omega_X/\Omega_{0}})({t_0/\tau_X})\sim5\times10^{-11},~
$
to account for the observed flux of UHECR events.
Here $t_0$ is the age of the universe, $\tau_X$ the lifetime
of the meta--stable state, $\Omega_{0}$ is the critical mass density
and $\Omega_{X}$ is the relic mass density of the meta--stable state.
It is evident that the parameters of the super--heavy meta--stable states
are sufficiently flexible to accommodate the observed flux of UHECR,
while evading other constraints \cite{cfp}.

Superstring theory inherently possesses the ingredients
that naturally give rise to super--heavy meta--stable states.
Such states arise in string theory due to the
breaking of the non--Abelian gauge symmetries by Wilson lines.
The massless spectrum then contains states with fractional electric
charge or ``fractional'' $U(1)_{Z^\prime}$ charge \cite{ww,eln,ccf}.
The lightest states are meta--stable due to a local gauge, or discrete,
symmetry \cite{ccf,lds}.
This phenomenon is of primary importance for superstring phenomenology.
The main consequence is that it generically results in
super--massive states that are meta--stable.
The super--heavy
states can then decay via the nonrenormalizable operators,
which are produced from exchange of heavy string modes,
with lifetime $\tau_x>10^{7-17}\;{\rm years}$ \cite{eln,ben,cfp}.
The typical mass scale of the exotic states will exceed the
energy range accessible to future collider experiments by several
orders of magnitude. The exotic states are rendered super--massive
by unsuppressed mass terms \cite{fcp}, or are confined
by a hidden sector gauge group \cite{eln}.
String models may naturally produce mass scales of the
required order, $M_X\approx10^{12-13}{\rm GeV}$,
due to the existence of an hidden sector that typically contains
non--Abelian $SU(n)$ or $SO(2n)$ group factors.
The hidden sector dynamics are set by the initial
conditions at the Planck scale,
and by the hidden sector gauge and matter content,
$M_X~\sim~\Lambda_{\rm hidden}^{\alpha_s,M_S}(N,n_f).$
Finally, the fact that $M_X\sim10^{12-13}{\rm GeV}$ implies
that the super--heavy relic is not produced in thermal
equilibrium and some other production mechanism is responsible
for generating the abundance of super--heavy relic \cite{ccf}.

The forthcoming cosmic rays observatories can therefore provide
fascinating experimental probes, both to the physics
above the electroweak scale as well as to more exotic
possibilities at a much higher scale.
It is therefore
imperative to develop the theoretical tools to decipher
the data from these experiments.
Moreover, improved information on the colliding
primaries may reveal important clues on the properties
of the decaying meta--stable state, which further motivates
the development of such techniques.
In this paper we make a
modest step in this direction, by studying possible modifications
of air shower simulations, that incorporate the possible
effects of new physics above the electroweak scale. This
is done by varying the cross section in the air shower
codes that are used by the experimentalists.
In this respect we assume here for concreteness
that the new physics above the electroweak scale remains
perturbative and preserve unitarity, as in the
case of supersymmetric extensions of the Standard Model.
This in turn is motivated by the success of supersymmetric
gauge coupling unification \cite{mssmunification} and their
natural incorporation in string theories. In the case of
supersymmetry the deviations from the Standard Model are
typically in the range of a few percent,
a quantitative indication which we take as our reference point for
study.

\subsection{Possible Developments}
Even if the forthcoming experiments
will confirm the existence of UHECR
events, it remains to be seen whether
any new physics can be inferred from the results. We
will argue that this is a very difficult question.

A possible way, in the top-down models of the UHECR
interaction is to optimize the analysis of any new high energy
primary interaction.
One should keep in mind that the information carried by the
primaries in these collisions
is strongly ``diluted'' by their interaction with the atmosphere and that
large statistical fluctuations are immediately generated both by the
randomness of the first impact, the variability
in the zenith angle of the impact, and the natural fluctuations in the
- extremely large - phase space available
at those energies. We are indeed dealing with {\em extreme} events.
These uncertainties are clearly
mirrored even in the existing Monte Carlo codes for the simulation of air
showers, and, of course,
in the real physical process that these complex Monte Carlo
implementations try, at their best,
to model (see also \cite{Sciutto} for the discussion of simulation issues) .
Part of our work will be concerned primarily  with trying to assess, by
going through extensive
air-shower simulations using existing interaction models - at the
GZK and comparable energies - the main features
of the showers, such as the multiplicities at various heights and
on the detector plane. We will
illustrate the geometry of a typical experimental
setup to clarify our method of analysis and
investigate in detail some geometrical observables.

A second part of our analysis will be centered
around the implications of a modified first impact on the multiplicities
of the subcomponents. Our analysis here is just
a first step in trying to see whether a modified first impact
cross section has any implication on the multiplicity structure of the
shower. The analysis is computationally very expensive
and has been carried out using a rather simple strategy
to render it possible. We critically comment on
our results, and suggest some possible improvements
for future studies.

\section{Simulation of Airshowers}
The quantification of the variability and parametric dependence
of the first impact in the formation of extensive airshowers can
be discussed, at the moment,
only using Monte Carlo event generators. Although various attempts
have been made
in the previous literature to model the spectrum of a
generic ``X-particle'' decay  in various
approximations, all of them include - at some
level and with variants - some new physics
in the generation of the original spectrum.
In practice what is seen at experimental level is
just a single event, initiated by a single hadron (a proton)
colliding with an air nucleus
(mostly of oxygen or nitrogen) within the 130 km depth of the
Earth atmosphere.
Our studies will show that
the typical strength of the interaction
of the primary at the beginning of the showers - at
least using the existing Monte Carlo codes -
has to increase fairly dramatically in order
to be able to see - at the experimental level -
any new physics.

Our objective here is to assess the actual possibility, if any, to
detect new physics from the
high energy impact of the primary cosmic ray assuming that other channels
open up at those energies.
Our investigation here is focused on the case of supersymmetry,
which is the more widely accepted extension of the Standard Model.
Other scenarios are left for future studies.

We recall that at the order of the GZK cutoff,
the center of mass energy of the first collision
reaches several hundreds of TeVs and is, therefore,
above any supersymmetric scale, according to current MSSM models.
It is therefore reasonable to ask weather supersymmetric interactions
are going to have any impact on some of the observables that are
going to be measured.

We will provide enough evidence that supersymmetric effects in
total hadronic cross sections cannot
raise the hadronic nucleon nucleon cross section
above a (nominal) 100\% upper limit. We will then
show that up to such limit the fluctuations in 1)
the multiplicity distributions of the most important
components of the (ground) detected airshowers and 2)
the geometric distributions of particles
on the detector are overwhelmly affected by natural (statistical)
fluctuations in the formation of the
air showers and insignificantly by any interaction whose strength
lays below such 100\% nominal limit.

In order to proceed with our analysis we need to define a set
of basic observables which can be used in the characterization of
the shower at various heights in the atmosphere.

There are some basic features of the shower that are important in
order to understand its structure and can be summarized in: 1)
measurements of its multiplicities in the main
components; 2) measurements of the geometry of the shower.
Of course there are obvious limitations in the study of the
development of the shower at the various levels,
since the main observations are carried out on the ground. However,
using both Cerenkov telescopes and fluorescence measurements by satellites
one hopes to reconstruct the actual shape of the shower as it develops
in the atmosphere.

To illustrate the procedure that we have
implemented in order to characterize the shower,
we have assumed that the first (random) impact of the incoming
primary (proton)
cosmic ray takes place at zero zenith angle, for simplicity. We have not
carried out simulations at variable zenith, since our objective is to
describe the
main features of the shower in a rather simple, but realistic, setting.
We have chosen a
flat model for the atmosphere and variable first impacts, at energies mainly
around the GZK cutoff region. Our analysis has been based on CORSIKA
\cite{CORSIKA}
and the hadronization model chosen has been QGSJET \cite{QGSJET}.

Measurements at any level are performed taking the arrival axis (z-axis)
of the primary as center of the detector. The geometry of the shower on
the ground and
at the various selected observation levels has been always measured
with respect to this
axis. The ``center'' of the detector is, in our simulations, assumed to be the
point at which the z-axis intersects the detector plane. Below,
the word ``center'' refers to this particular geometrical setting.

The shower develops according to an obvious cylindrical symmetry around the
vertical z-axis, near the center. The various components of the showers
are characterized at any observation level by this cylindrical symmetry.
Multiplicities are plotted after integration over the azimuthal angle and shown
as a function of the distance from the core (center), in the
sense specified above.

The showers show for each subcomponent specific locations of
the maxima and widths
of the associated distributions. We will plot the positions
of the maxima along the entire spatial extent of the shower in the
atmosphere. These plots are useful in order to have an idea of
what is the geometry of the
shower in the 130 km along which it develops.

\section{Features of the Simulation}
Most of our simulations are carried out at two main energies,
$10^{19}$ and $10^{20}$ eV. Simulations have been performed on a
small cluster running a communication protocol (openmoses)
which distributes automatically the computational load. The simulation
program follows each secondary from beginning to end and is extremely time
and memory intensive. Therefore,
in order to render our computation manageable we have
implemented in CORSIKA the thinning option \cite{thinning},
which allows to select only a
fraction of the entire shower and followed its development from
start to end.
We recall that CORSIKA is, currently, the main program used by the
experimental collaborations for the analysis of cosmic rays.
The results have been corrected statistically in order to reproduce
the result of the actual (complete) shower. The CORSIKA output has been
tokenized
and then analyzed using various intermediate software written by us.
The number of events generated, even with the thinning algorithm, is huge
at the GZK energy and requires an appropriate handling of the final data.
We have performed sets of run and binned the data using bins of 80 events,
where an event is a single impact with its given parametric dependence.
The memory cost of a statistically significant set of
simulations is approximately 700 GigaBytes, having selected in our
simulations a maximal number of
observations levels (9) along the entire height of the atmosphere.

\subsection{Multiplicities on the Ground}
\begin{figure}[tbh]
{\centering \subfigure[
]{\resizebox*{15cm}{!}{\rotatebox{-90}{\includegraphics{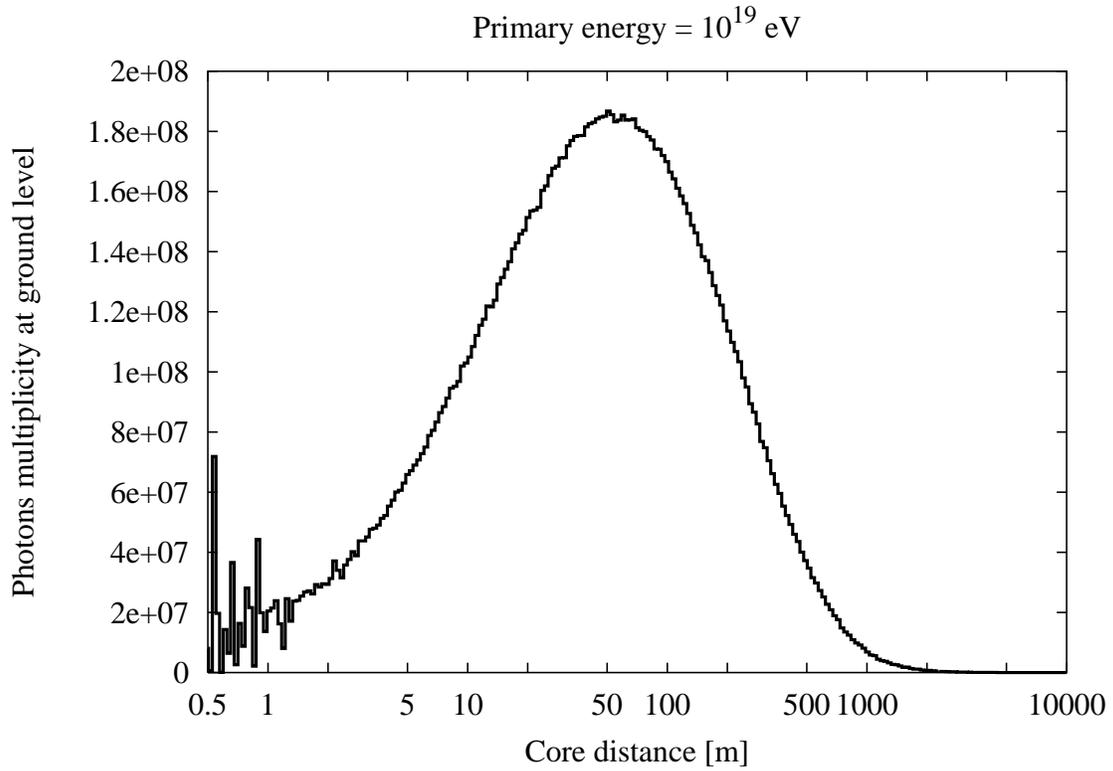}}}} \par}
{\centering \subfigure[
]{\resizebox*{13cm}{!}{\rotatebox{-90}{\includegraphics{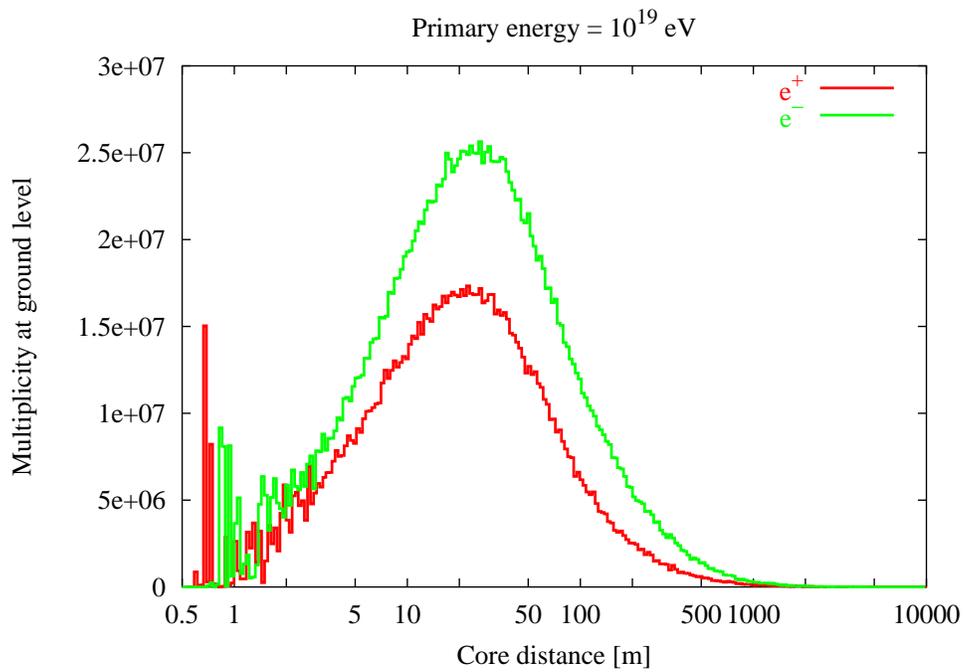}}}} \par}
\caption{Multiplicities of photons and \protect\( e^{\pm }\protect \)
at the ground level with a proton primary of \protect\( 10^{19}\protect \)
eV as a function of the distance from the core of the shower.}
\label{first}
\end{figure}

We show in Fig.~\ref{first} results for the
multiplicities of the photon component and of the $e^\pm$ components at
$10^{19}$ eV plotted against the distance from the core (center) of the
detector. For photons,
the maximum of the shower is around 90 meters from the center, as measured on
the plane of the detector. As evident from the plot,
the statistics is lower as we get closer to the central
axis (within the first 10
meters from the vertical axis), a feature which is typical of all
these distributions, given the
low multiplicities measured at small distances from the center.
For electrons and positrons
the maxima also lie within the first 100 meters, but slightly
closer to the center and are down
by a factor of 10 in multiplicities with
respect to the photons.
Positron distributions are
suppressed compared to
electron distributions. It is also easily noticed that the lower
tail of the photon distribution
is larger compared to the muon distribution, but all the
distributions show overall similar widths, about 1 km wide.

\begin{figure}[tbh]
{\centering \subfigure[
]{\resizebox*{15cm}{!}{\rotatebox{-90}{\includegraphics{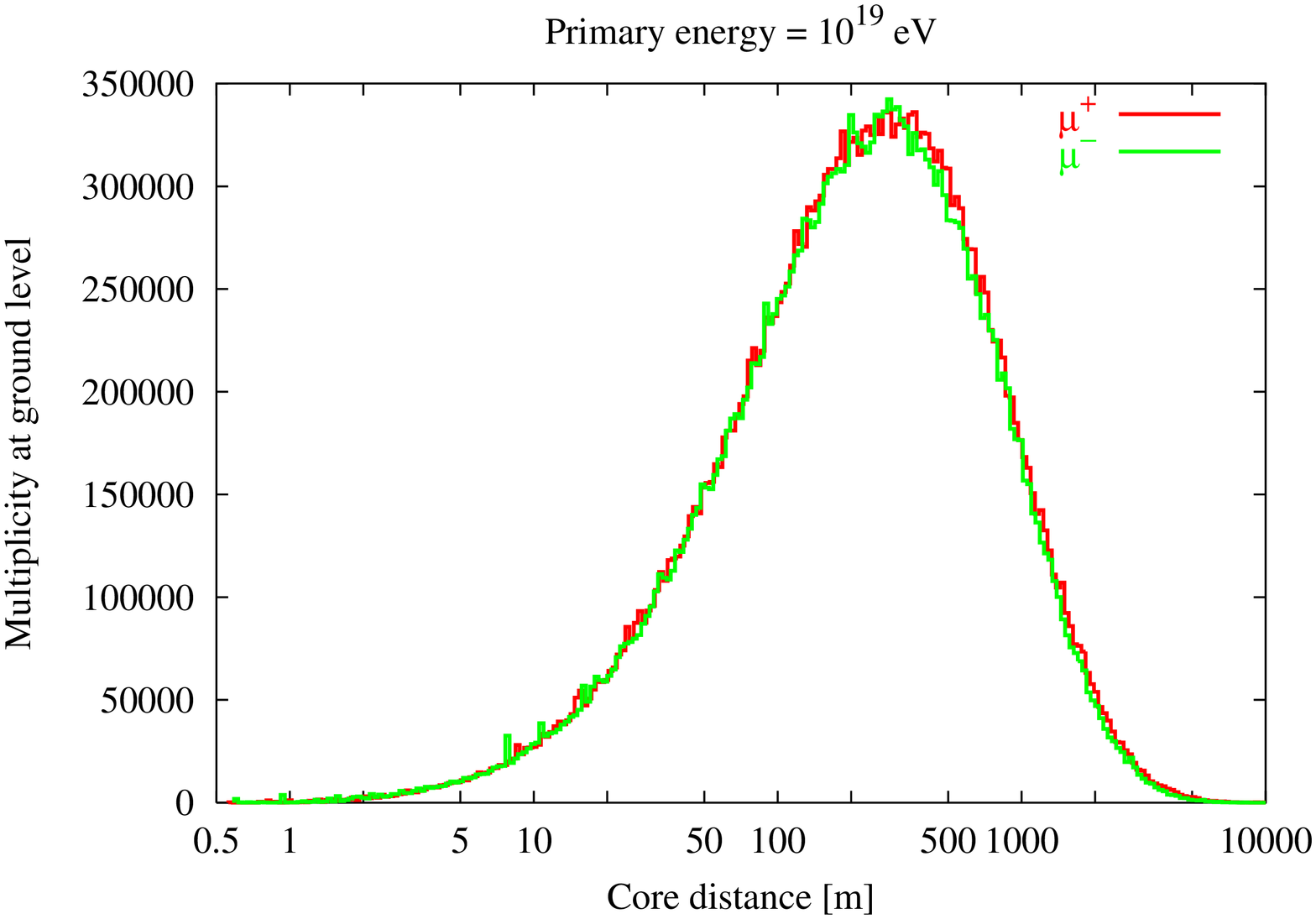}}}} \par}
\caption{Multiplicities of \protect\( \mu ^{\pm }\protect \)
at the ground level with a proton primary of \protect\( 10^{19}\protect \)
eV as a function of the distance from the core of the shower.}
\label{second}
\end{figure}

Multiplicities for muons
(see Fig.~\ref{second}) are a factor 1000 down with respect
to photons and 100 down with respect to electrons. The maxima of the muon
distributions are also at comparable distance as for the photons,
and both muon and antimuons
show the same multiplicity.

\begin{figure}[tbh]
{\centering \subfigure[
]{\resizebox*{13cm}{!}{\rotatebox{-90}{\includegraphics{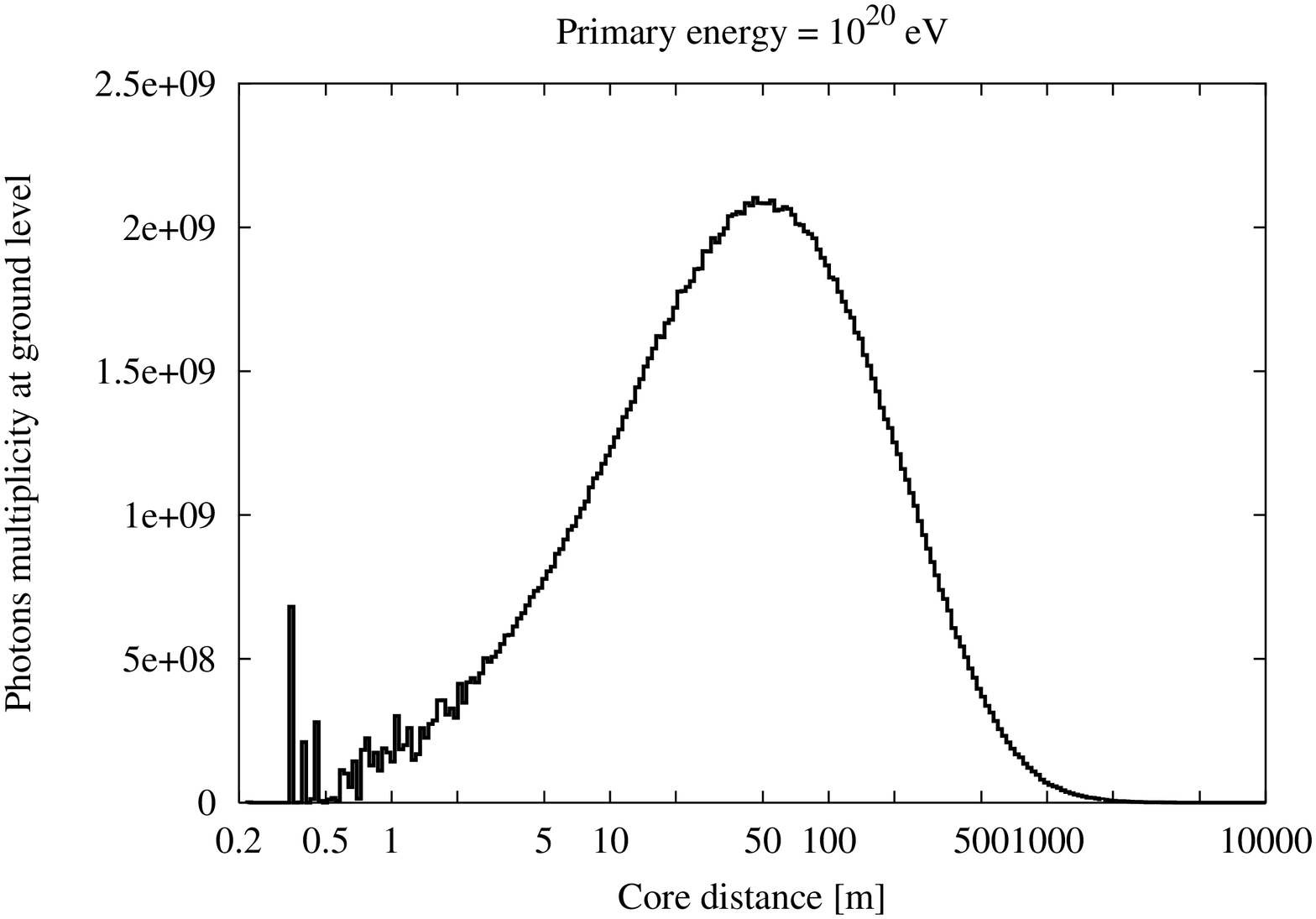}}}} \par}
{\centering \subfigure[
]{\resizebox*{15cm}{!}{\rotatebox{-90}{\includegraphics{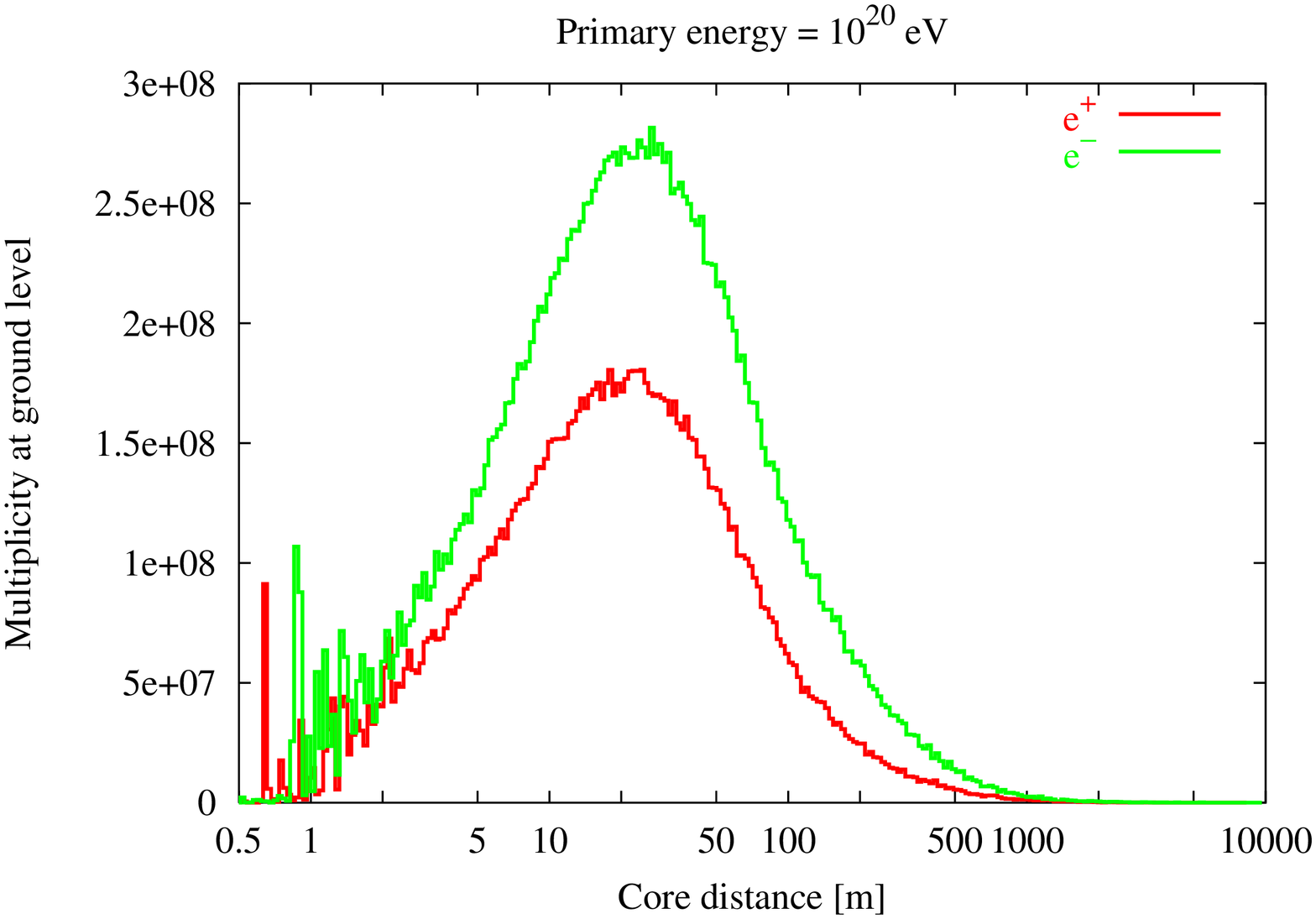}}}} \par}
\caption{Multiplicities of photons and \protect\( e^{\pm }\protect \)
at the ground level with a proton primary of \protect\( 10^{20}\protect \)
eV as a function of the distance from the core of the shower.}
\label{third}
\end{figure}

\begin{figure}[tbh]
{\centering \subfigure[
]{\resizebox*{15cm}{!}{\rotatebox{-90}{\includegraphics{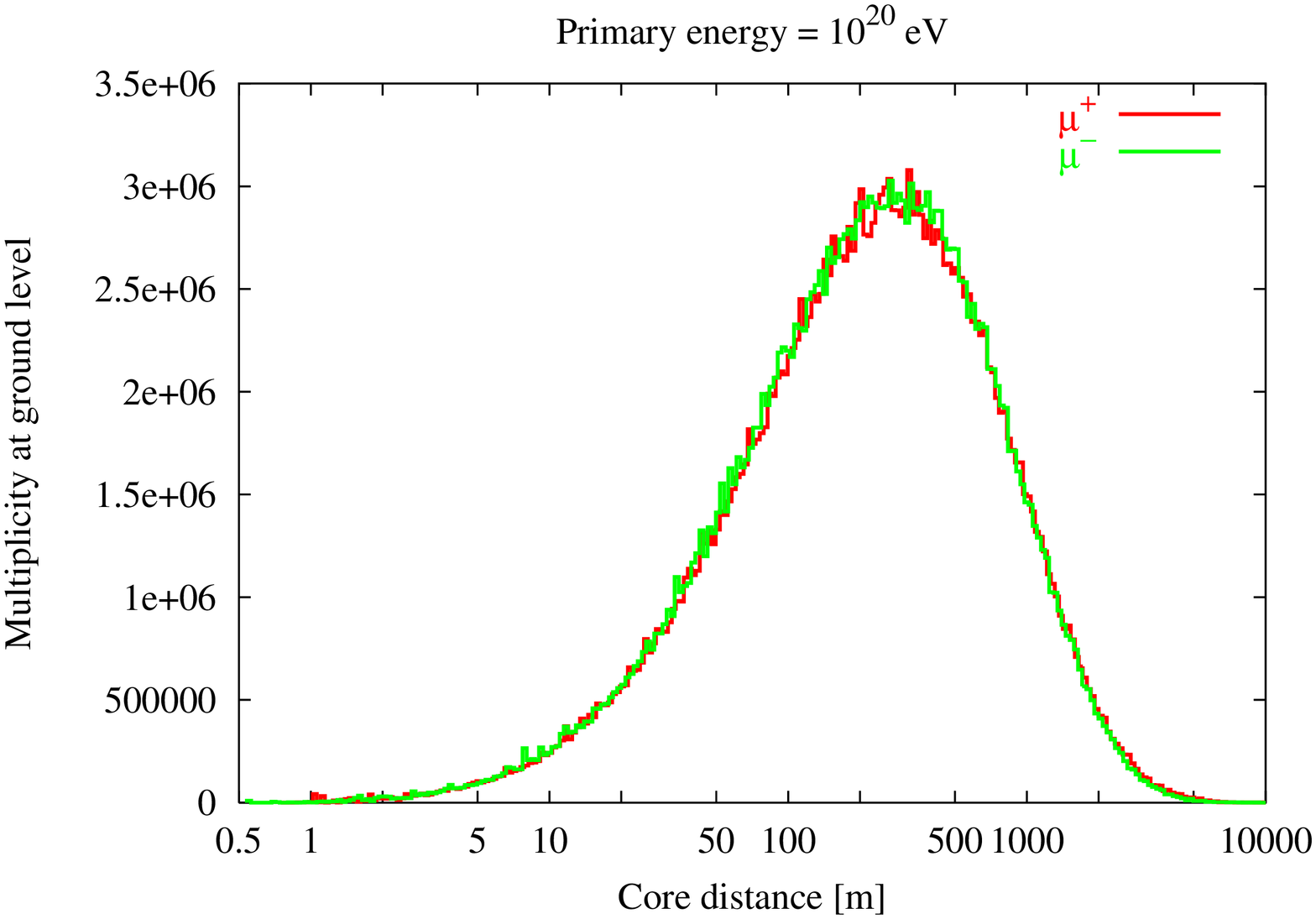}}}} \par}
\caption{Multiplicities of \protect\( \mu ^{\pm }\protect \)
at the ground level with a proton primary of \protect\( 10^{20}\protect \)
eV as a function of the distance from the core of the shower.}
\label{fourth}
\end{figure}

At the GZK energy (Figs \ref{third}, \ref{fourth})
the characteristics of the distributions of
the three main components (photons, electrons, muons)
do not seem to vary appreciably, except for the values of the multiplicities,
all increased by a factor of 10 respect to the previous plots.
The maxima of the
photons multiplicities are pushed away from the center,
together with their tails.
There appears also to be an increased
separation in the size of the multiplicities of electrons and positrons and a
slightly smaller width for the photon distribution compare to the lower energy
result ($10^{19}$ eV). We should mention
that all these gross features of the showers can possibly be tested after
a long run time of the experiment.
Our distributions have been obtained averaging over sets of 80
events with independent first impacts.

\section{Missing multiplicities?}
\begin{figure}[tbh]
{\centering
\resizebox*{15cm}{!}{\rotatebox{-90}{\includegraphics{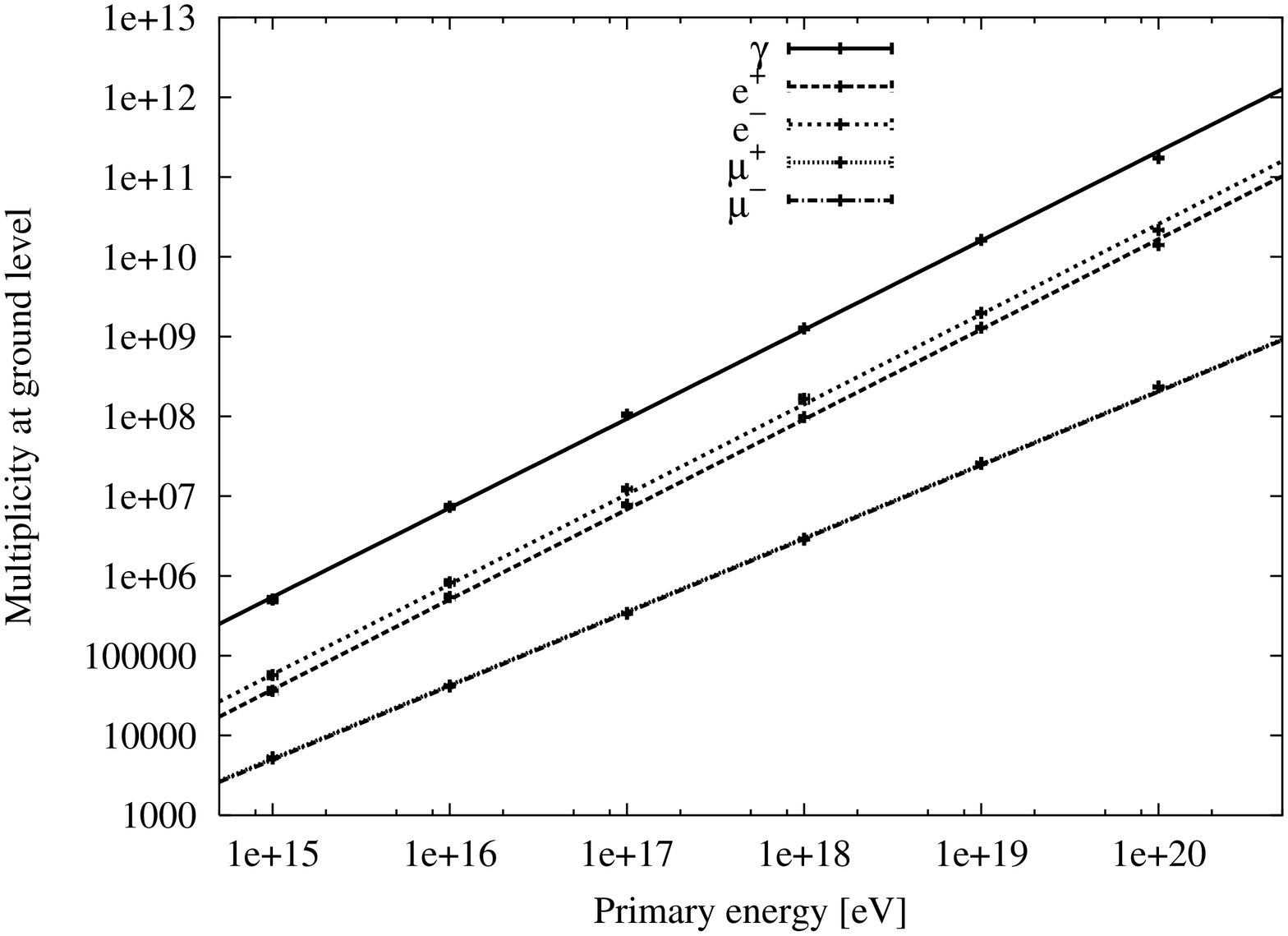}}} \par}
\caption{Multiplicities of photons, \protect\( e^{\pm }\protect \), \protect\(
\mu ^{\pm }\protect \)
at the ground level as a function of the primary energy. Due to the
logarithmic scale, \protect\( \mu ^{+}\protect \) and \protect\( \mu
^{-}\protect \)
look superimposed.}
\label{fifth}
{\centering
\resizebox*{15cm}{!}{\rotatebox{-90}{\includegraphics{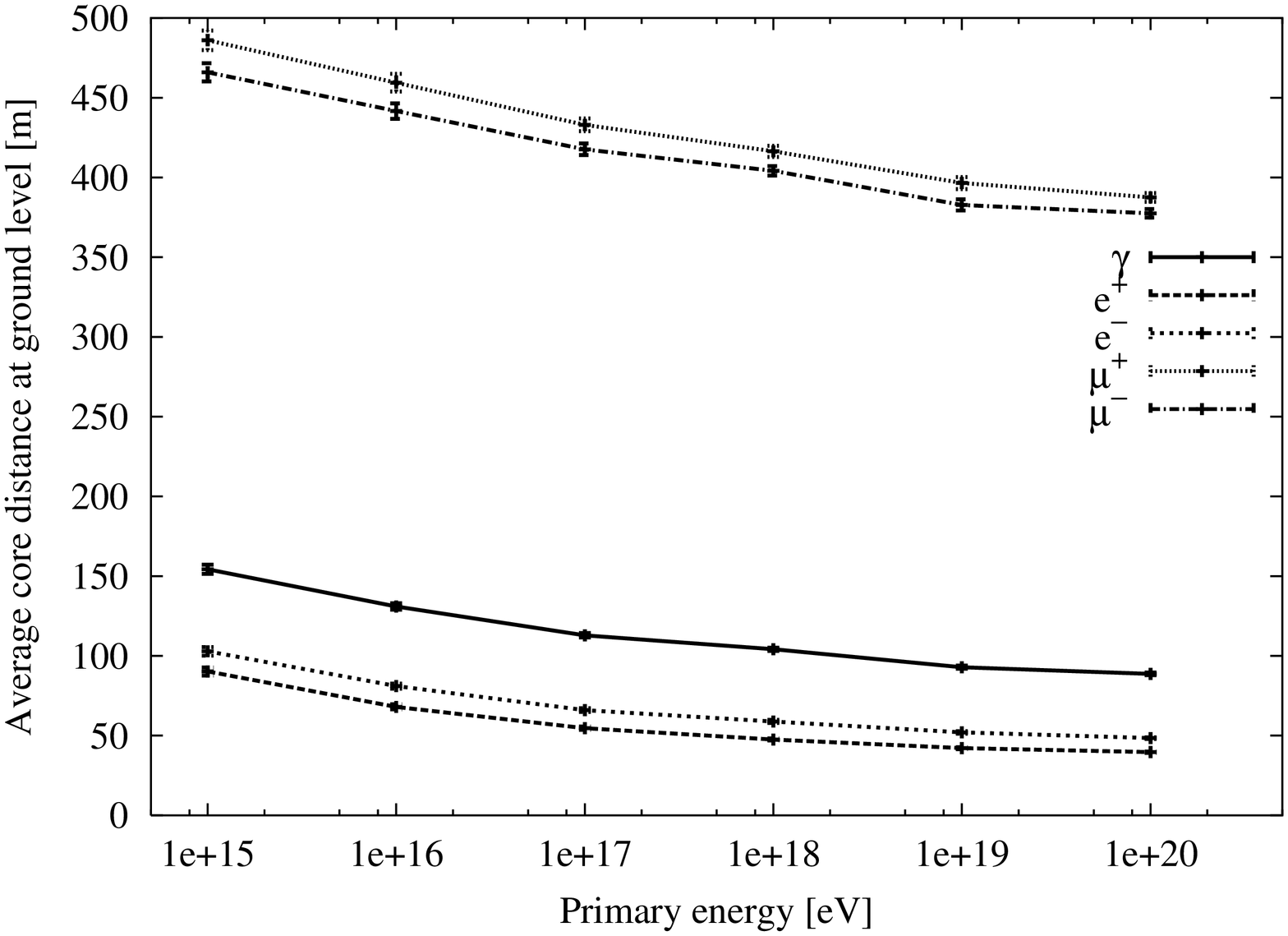}}} \par}
\caption{Average distance from the core of the shower of photons, \protect\(
e^{\pm }\protect \),
\protect\( \mu ^{\pm }\protect \) at the ground level as a function
of the primary energy.}
\label{sixth}
\end{figure}

Other inclusive observables which are worth studying are the total
multiplicities, as measured
at ground level, versus the total energy of the primary. We show in
Fig.~\ref{fifth} a double logarithmic plot of the total multiplicities
of the various components versus the primary energy in the range
$10^{15}-10^{20}$ eV, which appears to be strikingly linear.
{}From our result it appears that the multiplicities can be
fitted by a relation of the form  y=m*x+q,
where  $y={\rm Log(N)}$ and $x= {\rm Log(E)}$
or $N(E)= 10^q  x^m$. The values of $m$ and  $q$ are given by

\beqa
\gamma : && m = 1.117 \pm 0.011;\,\,\, q = -11.02 \pm 0.19\nonumber \\
e^+: && m = 1.129 \pm 0.011; \,\,\,    q = -12.36 \pm 0.18 \nonumber \\
e^-: && m = 1.129 \pm 0.012;  \,\,\,   q = -12.17 \pm 0.20 \nonumber\\
\mu^+: && m = 0.922 \pm 0.006; \,\,\,  q = -10.12 \pm 0.10 \nonumber \\
\mu^- && m = 0.923 \pm 0.006;  \,\,\,  q = -10.15 \pm 0.09 \nonumber \\
\eeqa
where $m$ is the slope. $m$ appears to be almost universal for all the
components,
while the intercept $(q)$ depends on the component.
The photon component is clearly dominant, followed by the electron, positron
and the
two muon components which appear to be superimposed.
It would be interesting to see whether missing energy effects, due, for
instance, to
an increased multiplicity rate toward the production of weakly
interacting particles can modify this type of inclusive measurements,
thereby predicting variations in the slopes of the multiplicities  respect
to the Monte Carlo predictions reported here. One could entertain
the possibility that a failure
to reproduce this linear behaviour could be a serious problem for the theory
and a possible signal of new physics. Given the large sets
of simulations that we have performed, the statistical
errors on the Monte Carlo results are quite small, and the Monte Carlo
prediction
appear to be rather robust. The difficulty of these measurements however,
lay mainly in the energy reconstruction of the primary, with the possibility of
a
systematic error. However, once the reconstruction of the energy of the primary
is under control among the various UHECR detectors, with a global calibration,
these measurements could be a possible test for new physics. At the moment,
however, we still do not have a quantification of the deviation
from this behaviour such as that induced by supersymmetry or other
competing theoretical models.

\subsection{Directionality of the Bulk of the Shower}
Another geometrical feature of the shower is the position
of its bulk
(measured as the opening of the radial cone at the radial distance
where the maximum is achieved)
as a function of the energy. This feature is illustrated in Fig.~\ref{sixth}.
Here we have plotted the averaged location
of the multiplicity distribution as a function of the energy of the incoming
primary. The geometrical center of the distributions
tend to move slightly toward the vertical axis (higher collimation) as the
energy
increases. From the same plot it appears that
the distributions of electrons, positrons and photons
are closer to the center of the detector compared to the muon-antimuon
distributions. As shown in the figure, the
statistical errors on these results appear to be rather small.

\subsection{The Overall Geometry of the Shower}
As the shower develops in the atmosphere, we can monitor
both the multiplicities of the various components and the average
location of the bulk of the distributions at various observation levels.
As we have mentioned above, we choose up to 9 observation
levels spaces at about 13 km from one another. The lowest observation
level is, according to our conventions, taken to coincide
with the plane of the detector.

\begin{figure}[tbh]
{\centering \subfigure[
]{\resizebox*{15cm}{!}{\rotatebox{-90}{\includegraphics{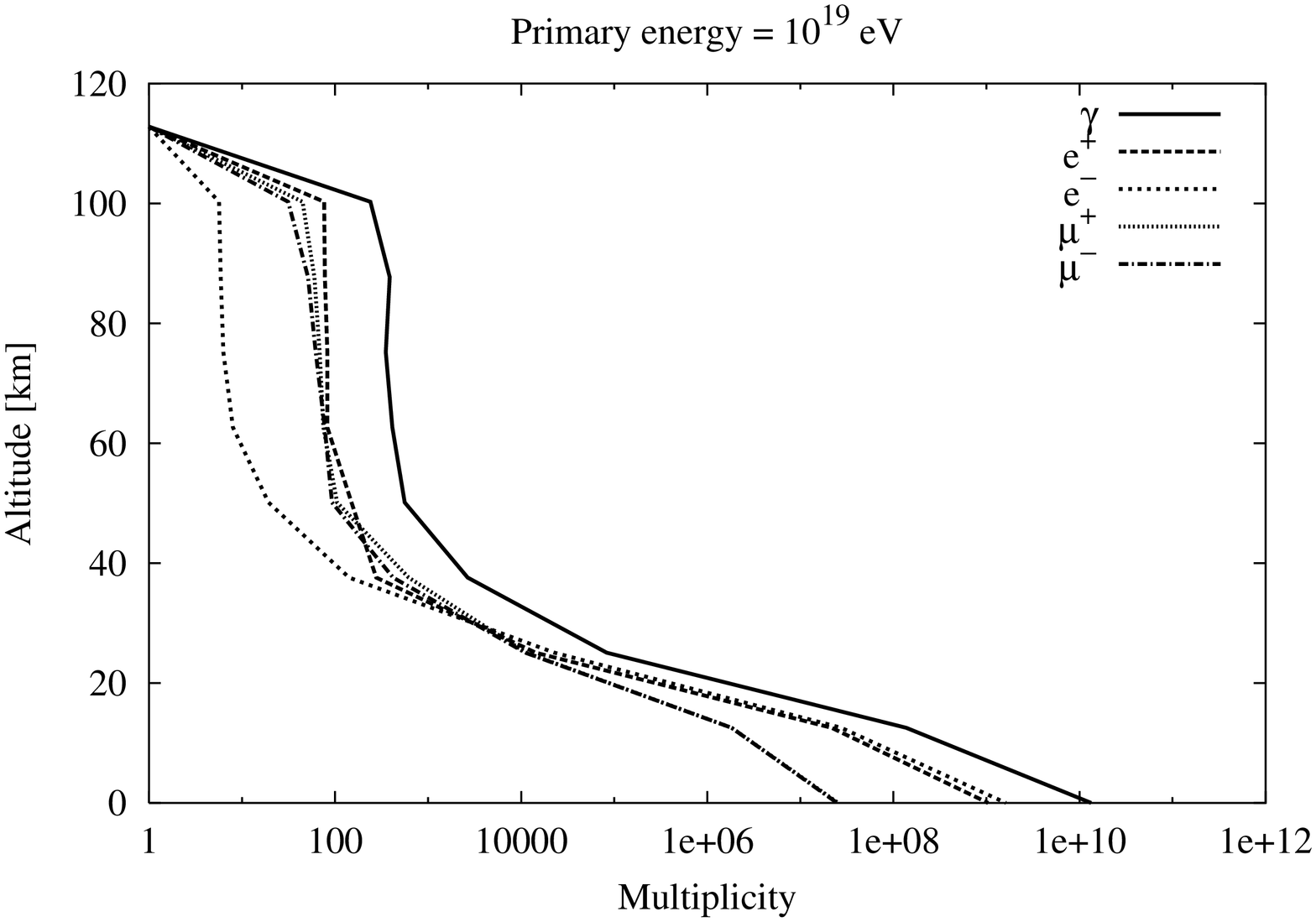}}}}
\par}
{\centering \subfigure[
]{\resizebox*{15cm}{!}{\rotatebox{-90}{\includegraphics{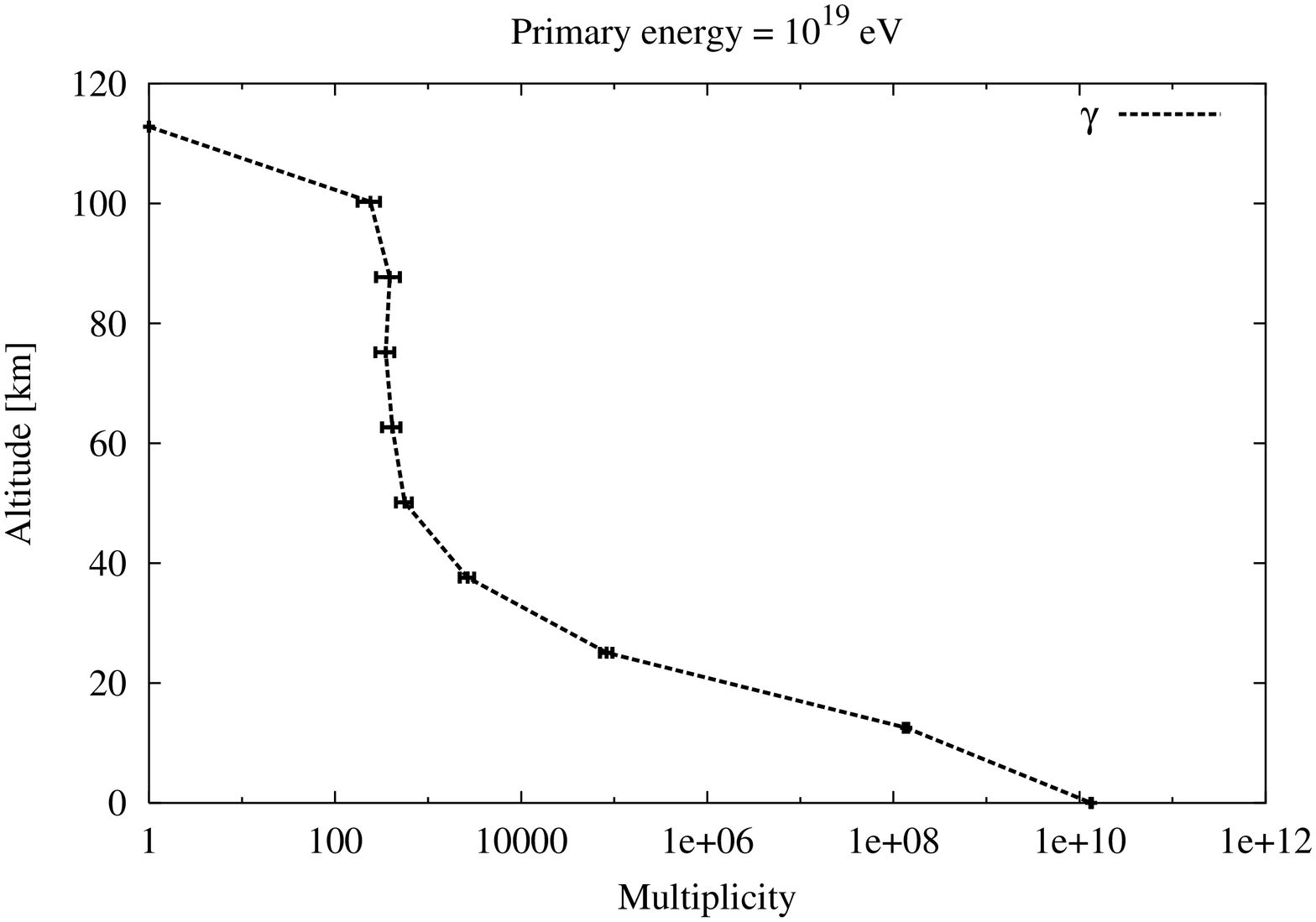}}}}
\par}
\caption{Multiplicities of photons, \protect\( e^{\pm }\protect \), \protect\(
\mu ^{\pm }\protect \)
at various levels of observations for a primary energy of \protect\(
10^{19}\protect \)
eV. The first impact is forced to occur at the top of the atmosphere.
In the subfigure (b) we show the uncertainties just in the case of
the photons.}
\label{seventh}
\end{figure}

We show in Fig.~\ref{seventh}a a complete simulation of the shower using a set
of 9
observation levels, as explained above, at an energy of $10^{19}$ eV.
In the simulation we assume that a first impact occurs near the top of the
atmosphere
at an eight of 113 km and we have kept this first impact fixed.
The multiplicities
show for all the components a rather fast growth within the first 10 km of
crossing of the atmosphere after the impact,
with the photon components growing faster compared to the others.
The electron component also grows rather fast, and a similar behaviour is
noticed for the muon/antimuon components, which show a linear growth
in a logarithmic scale (power growth). In the following 40 km downward,
from a height of 100 km down to 60 km, all the components largely conserve
their
multiplicities. Processes of regeneration of the various components and their
absorption seem to balance. For the next 20 km, from a height of 60 km down to
40 km, all the components starts to grow,
with the photon component showing a faster (power growth) with the traversed
altitude. Slightly below 40 km of altitude the multiplicities of three
components seem to merge
(muons and electrons), while the photon component is still dominant
by a factor of 10 compared to the others. The final development of the
airshower is characterized by a drastic growth
of all the components, with a final reduced muon component, a larger electron
component and a dominant photon component. The growth
in this last region (20 km wide) and in the first 20 km after
the impact of the primary -in the upper part of the atmosphere- appear to be
comparable.
The fluctuations in the multiplicities of the components
are rather small at all levels, as shown (for the photon case) in
Fig.\ref{seventh}b.

\begin{figure}[tbh]
{\centering \subfigure[
]{\resizebox*{15cm}{!}{\rotatebox{-90}{\includegraphics{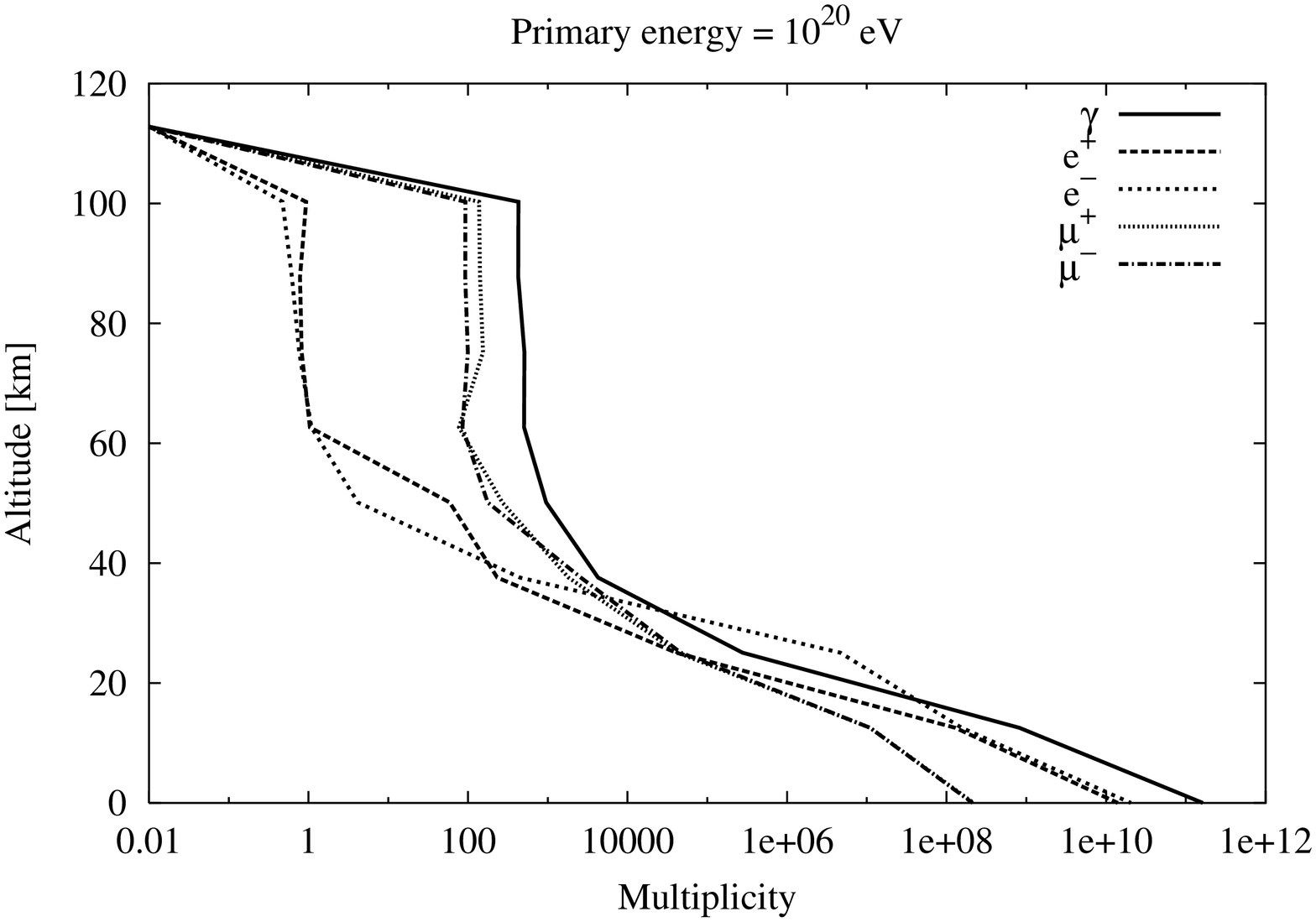}}}}
\par}

{\centering \subfigure[
]{\resizebox*{15cm}{!}{\rotatebox{-90}{\includegraphics{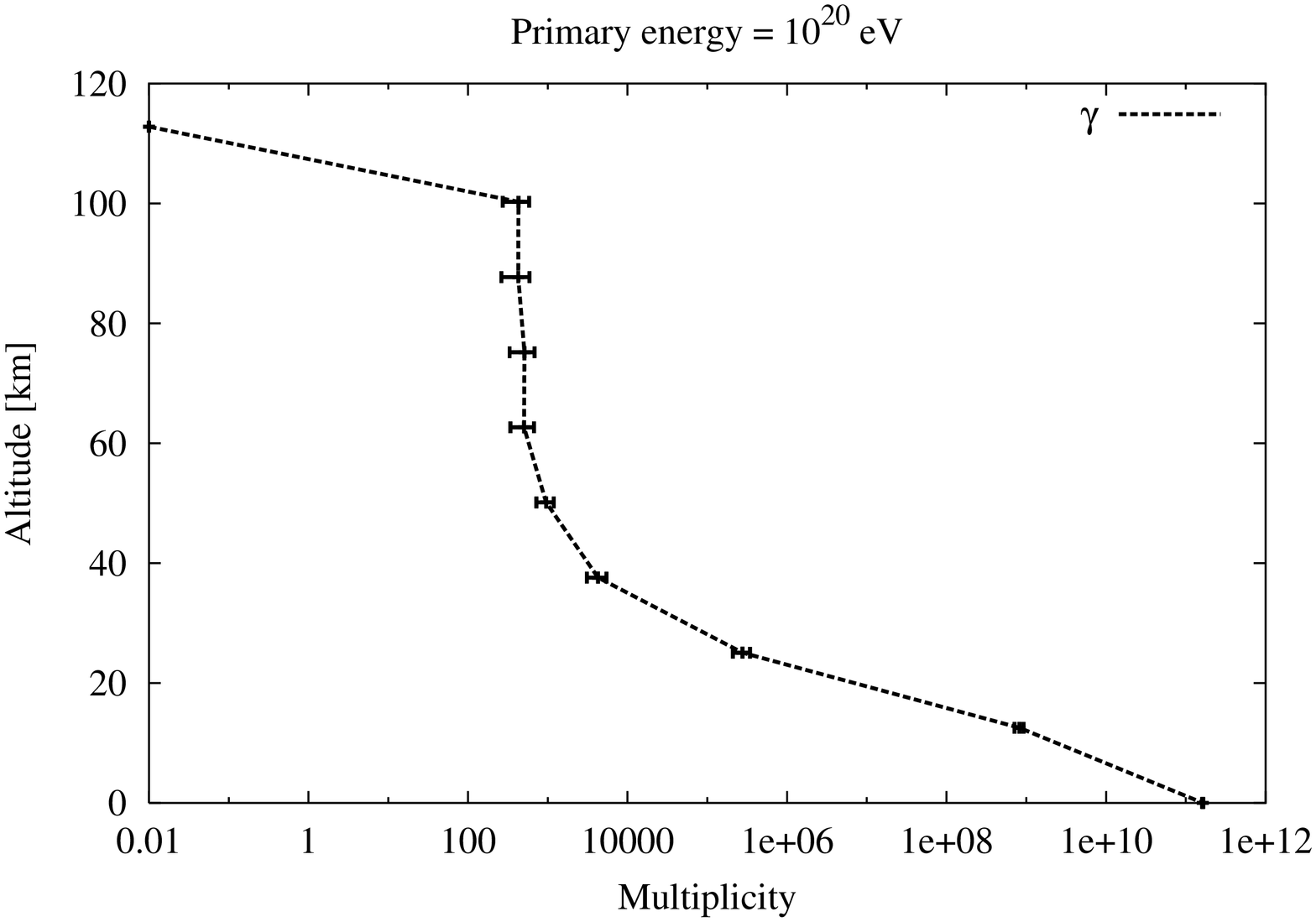}}}}
\par}
\caption{Multiplicities of photons, \protect\( e^{\pm }\protect \), \protect\(
\mu ^{\pm }\protect \)
at various levels of observations for a primary energy of \protect\(
10^{20}\protect \)
eV. The first impact is forced to occur at the top of the atmosphere.
In the subfigure (b) we show the uncertainties just in the case of
the photons.}
\label{eigth}
\end{figure}

As we reach the GZK cutoff, increasing the energy of the primary
by a factor of 10, the pattern just discussed in Fig.~\ref{seventh}a is
reproposed in Fig.~\ref{eigth}a, though -in this case- the growth of the
multiplicities
of the subcomponents in the first 20 km from the impact and in the last 20 km
is much stronger. The electron and the muon components appear
to be widely separated, while the electron and positron components
tend to be more overlapped. To the region of the first impact
-and subsequent growth- follows an intermediate region, exactly
as in the previous plots, where the two phenomena of production and absorption
approximately balance one another and the multiplicities undergo minor
variations. The final
growth of all the components is, a this energy, slightly anticipated compared
to
Fig. \ref{seventh}a, and starts to take place at a height of 40 km and above
and continues
steadily until the first observation level. The photon remains the
dominant component, followed by the electron and the muon component. Also
in this case the fluctuations (pictured for the photon component only, Fig.
\ref{eigth}b)
are rather small.

\subsection{The Opening of the Shower}
\begin{figure}[tbh]
{\centering \subfigure[
]{\resizebox*{15cm}{!}{\rotatebox{-90}{\includegraphics{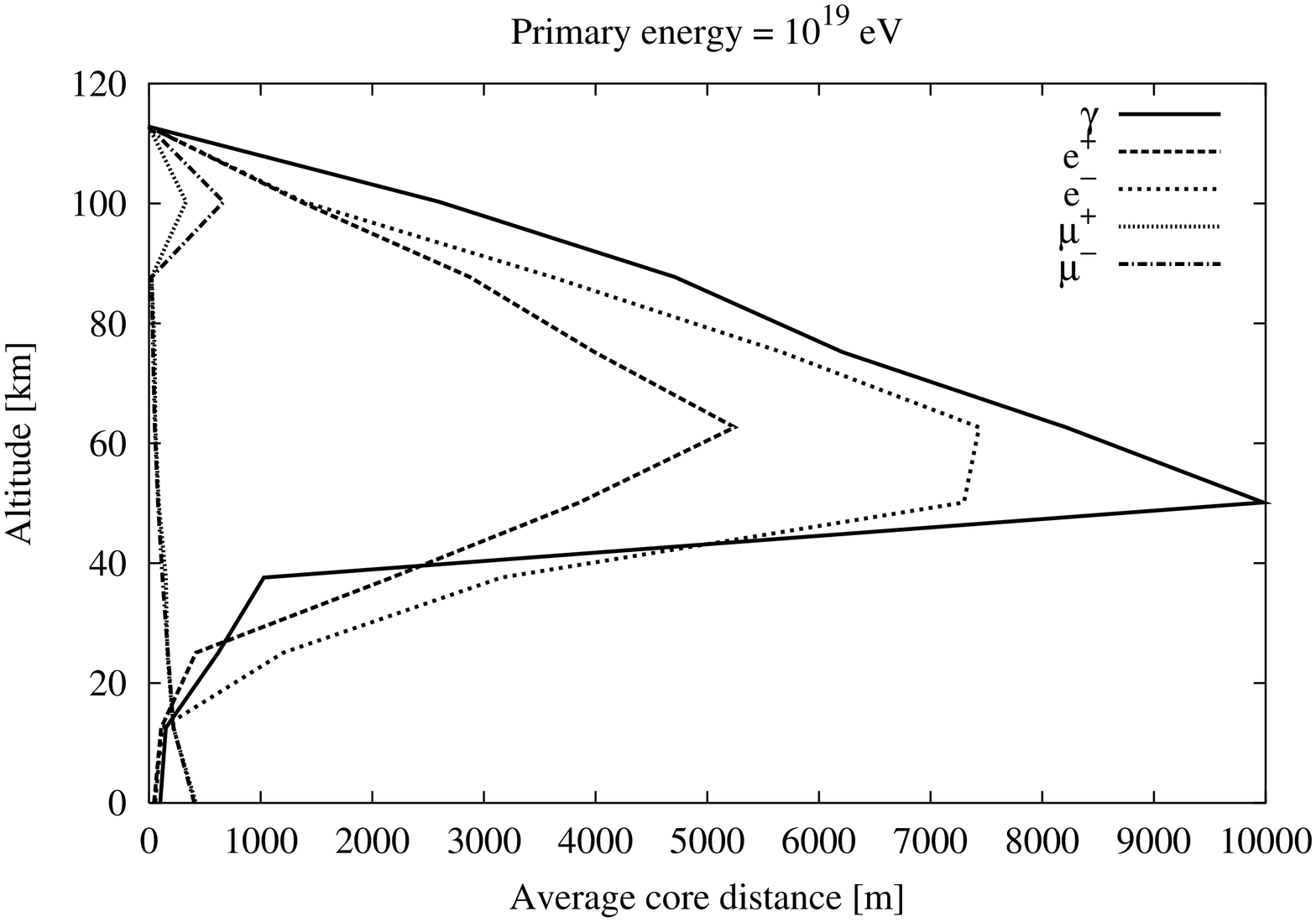}}}} \par}
{\centering \subfigure[
]{\resizebox*{15cm}{!}{\rotatebox{-90}{\includegraphics{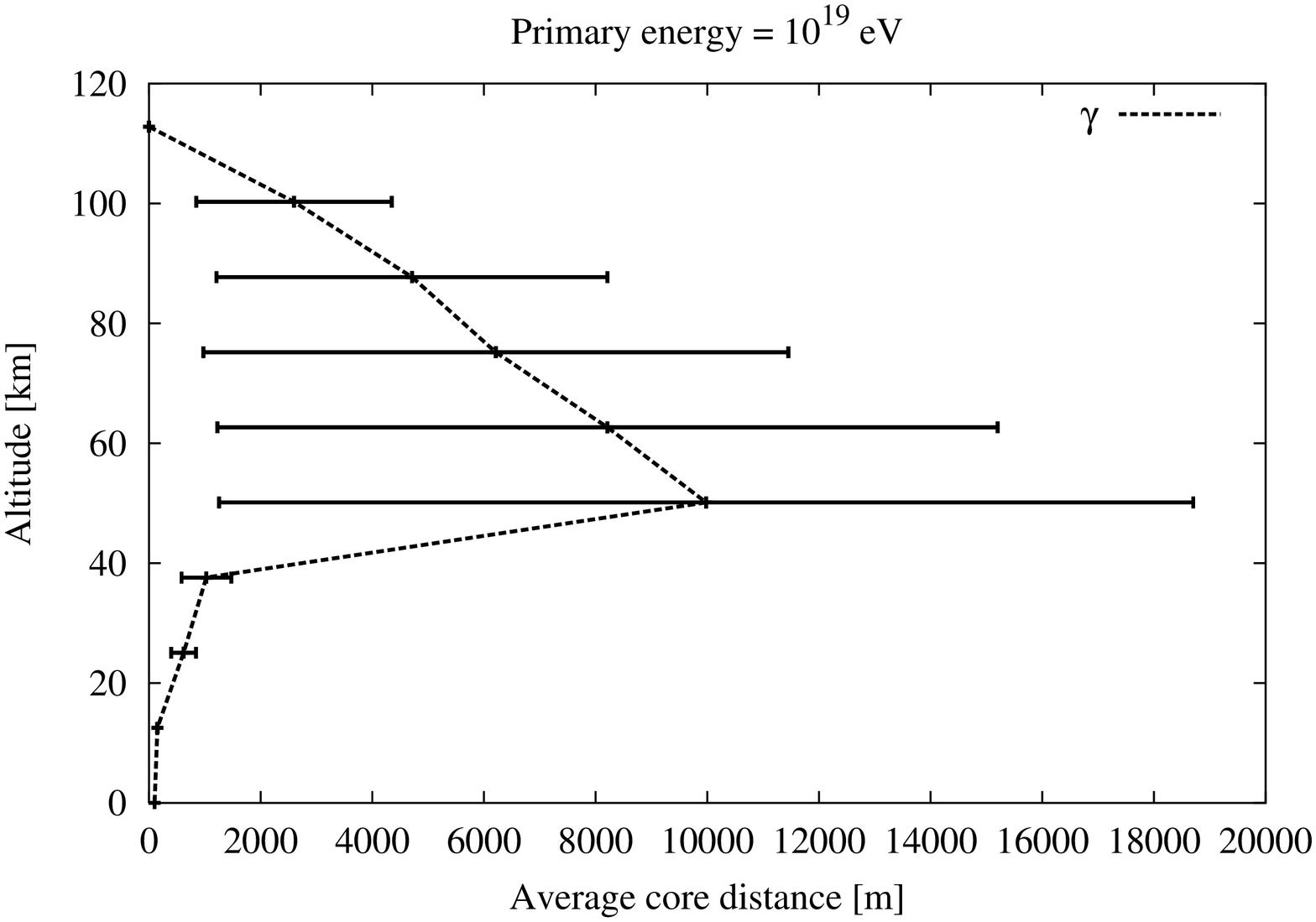}}}}
\par}
\caption{Average core distances of photons, \protect\( e^{\pm }\protect \),
\protect\( \mu ^{\pm }\protect \) at various levels of observations
for a primary energy of \protect\( 10^{19}\protect \) eV. The first
impact is forced to occur at the top of the atmosphere. In the subfigure
(b) we show the uncertainties just in the case of the photons.}
\label{ninth}
\end{figure}

\begin{figure}[tbh]
{\centering \subfigure[
]{\resizebox*{15cm}{!}{\rotatebox{-90}{\includegraphics{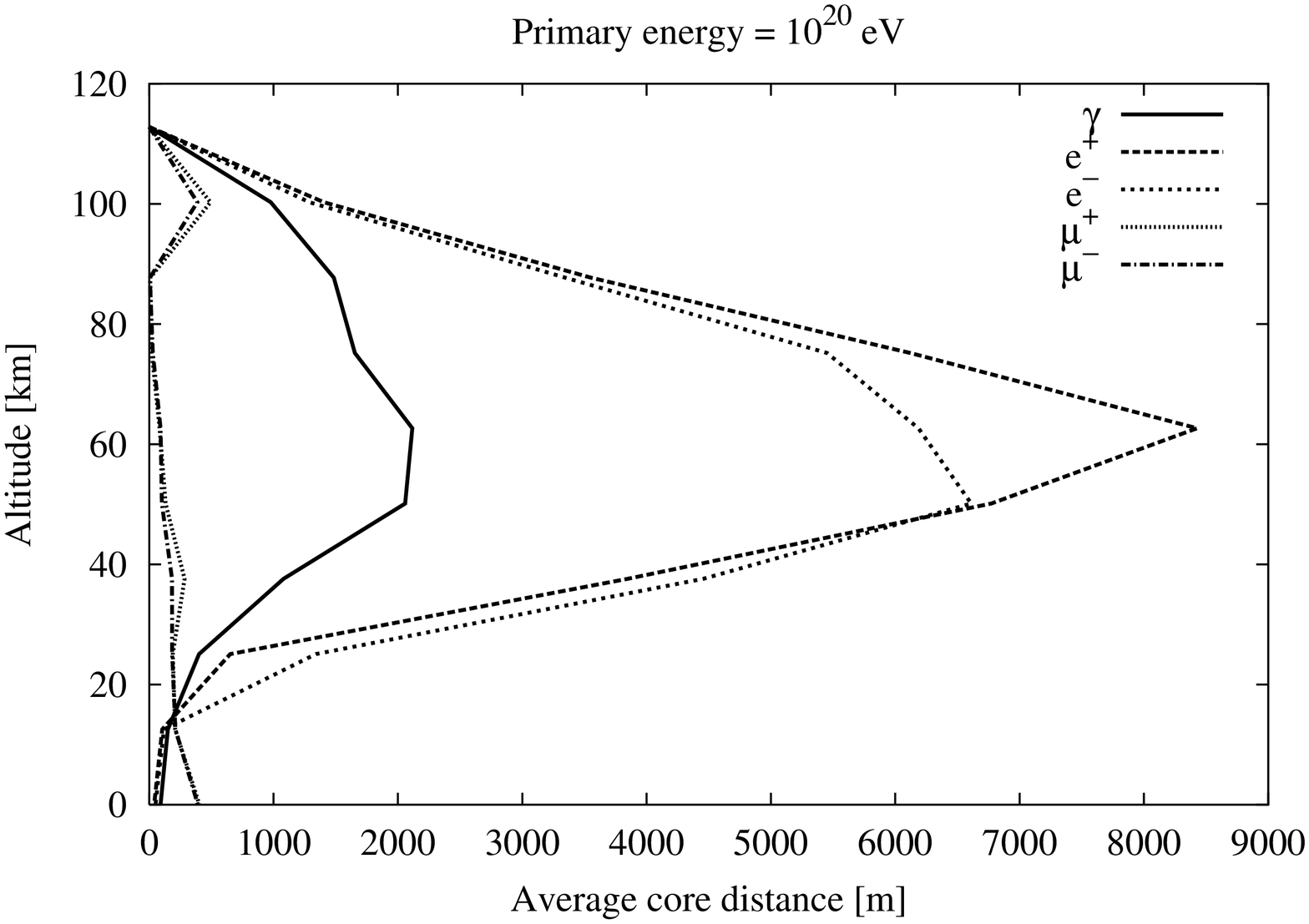}}}} \par}
{\centering \subfigure[
]{\resizebox*{15cm}{!}{\rotatebox{-90}{\includegraphics{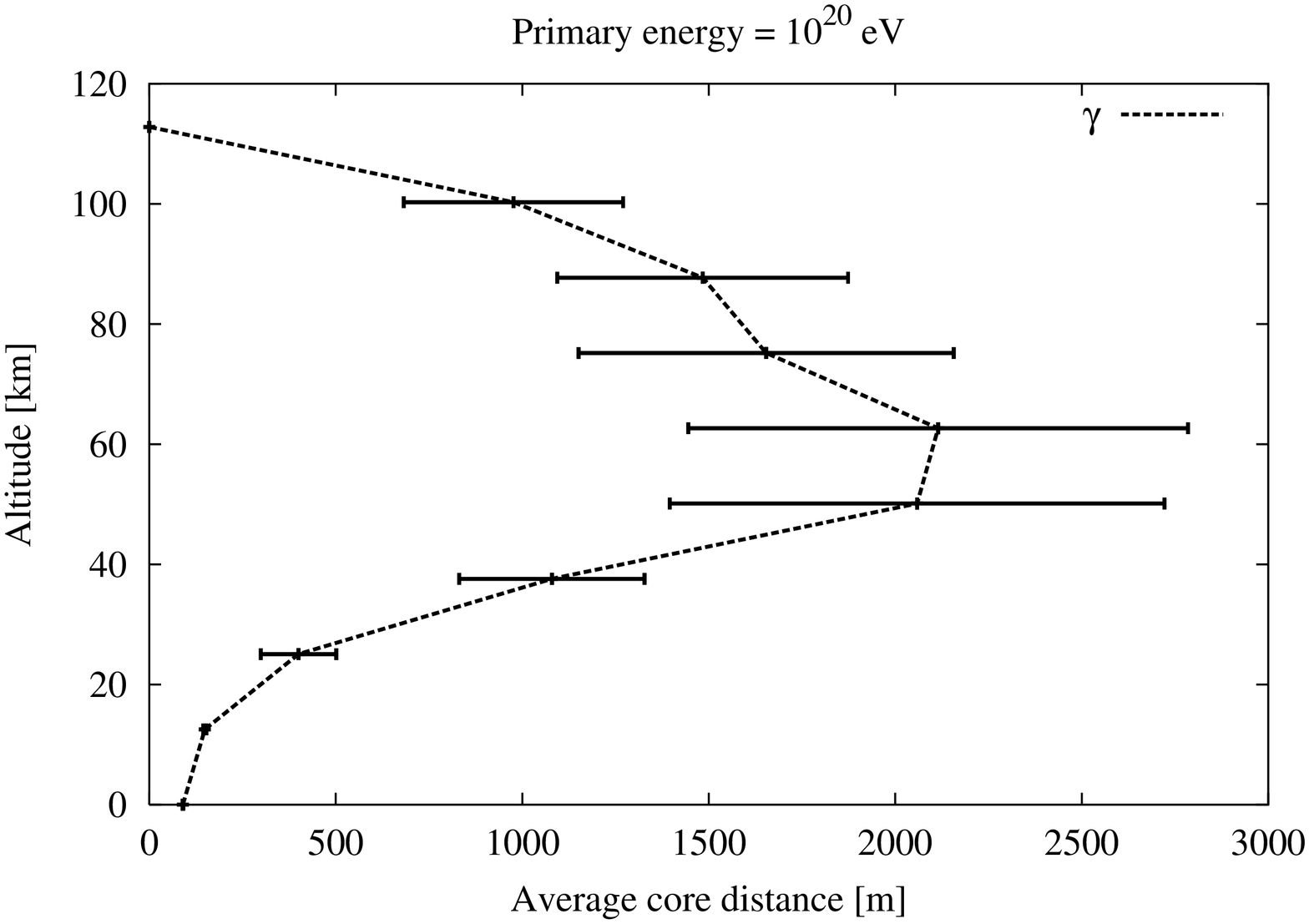}}}}
\par}
\caption{Average core distances of photons, \protect\( e^{\pm }\protect \),
\protect\( \mu ^{\pm }\protect \) at various levels of observations
for a primary energy of \protect\( 10^{20}\protect \) eV. The first
impact is forced to occur at the top of the atmosphere. In the subfigure
(b) we show the uncertainties just in the case of the photons.}
\label{tenth}
\end{figure}

In our numerical study the geometrical center of each component of the shower
is identified through
a simple average with respect to all the distances from the core
\beq
R_M\equiv=\frac{1}{N}\sum_i R_i
\eeq
where $N$ is the total multiplicity at each selected observation level, $R_i$
is the position
of the produced particle along the shower and $i$ runs over the single events.
This analysis has been carried out for 9 equally
spaced levels and the result of this study are shown in
Figs.~\ref{ninth},\ref{tenth} and
\ref{eigth} at two different values of energy ($10^{19}$ and $10^{20}$ eV). The
opening of the various components are clearly identified by these plots. We
start from Fig.~\ref{ninth}.
We have taken in this figure an original point of impact at a height of 113 Km,
as in the previous simulations.
It is evident that the photon component of the shower tends to spread rather
far and within
the first 20 km of depth into the atmosphere
has already reached an extension of about 2 km;
reaching a lateral extension of
10 km within the first 60 km of crossing of the atmosphere.

Starting from a height of 50 km down to 10 km,
the shower gets reabsorbed (turns toward the center) and is characterized
by a final impact which lays rather close to the vertical axis. Electrons
and photons follow a similar behaviour, except that for electrons whose
lateral distribution in the first section of the development of the shower is
more reduced.
The muonic
(antimuons) subcomponents appear to have a rather small opening and develop
mostly along the vertical axis of impact. In the last section of the shower
all the components get aligned near the vertical axis and hit the detector
within 1 km.

Few words should be said about the fluctuations.
At $10^{19}$ eV, as shown in Fig.~\ref{ninth}b, the fluctuations are rather
large,
especially in the first part of the development of the shower. These turn
out to be more pronounced for photons, whose multiplicity growth is large
and very broad. As we increase the energy of the primary to $10^{20}$ eV the
fluctuations in the lateral distributions (see Fig.~\ref{tenth}b) are overall
reduced, while the lateral distributions of the photons appear to be
drastically reduced (Fig.~\ref{tenth}a).

\section{Can we detect new physics at Auger?}
There are various issues that can be addressed,
both at theoretical and at experimental level, on this point,
one of them being an eventual confirmation
of the real existence of events above the cutoff.
However, even if these measurement will confirm their existence, it
remains yet to be seen whether any additional new physics can be inferred
just from an analysis of the air shower.
A possibility might be supersymmetry or any new underlying interaction,
given the large energy available in the first impact.
We recall that the spectrum of the decaying X-particle
(whatever its origin may be),
prior to the atmospheric impact of the
UHECR is of secondary relevance, since
the impact is always due to a single proton.
Unless correlations are found among different events - and by
this we mean that
a large number of events should be initiated by special types of primaries -
we tend to believe that effects due to new interactions are likely to
play a minor role.

In previous works we have analyzed in great detail the
effects of supersymmetry in the formation
of the hadronic showers. These studies, from our previous experience
\cite{CC,CC1,fragfun},
appear to be rather complex since they involve several possible
intermediate and large
final scales and cannot possibly be conclusive.
There are some obvious doubts
that can be raised over these analysis, especially when the DGLAP
equations come
to be extrapolated to such large evolution scales,
even with a partial resummation
of the small-x logarithms. In many cases results obtained in this
area of research by extrapolating results from collider phenomenology
to extremely
high energies should be taken with extreme caution in order to reduce
the chances of inappropriate hasty conclusions.
What is generally true in a first approximation is that supersymmetric effects
do appear to be mild \cite{CC,CC1,fragfun}. Rearrangements in the fragmentation
spectra or supersymmetric effects in initial state scaling violations are down
at the few percent level. We should mention that the generation of
supersymmetric
scaling violations in parton distributions, here considered to be the bulk of
the supersymmetric
contributions,
are rather mild if the entrance into the SUSY region takes
place ``radiatively''
as first proposed in \cite{CC,CC1}. This last picture might change in
favour of a more
substantial signal if threshold enhancements are also included in the
evolution,
however this and other related points have not yet been analyzed
in the current literature.

\subsection{The Primary Impact and a Simple Test}
Our objective, at this point, is to describe the structural properties
of the shower with an emphasis on the dynamics of the first impact of
the primary with the
atmosphere, and at this stage one may decide to look for the emergence of
possible new interactions, the most popular one being supersymmetry.

One important point to keep into consideration is
that the new physical signal carried by the primaries in these collisions
is strongly ``diluted'' by their interaction with the atmosphere and that
large
statistical fluctuations are immediately generated both by the
randomness of the first impact, the variability
in the zenith angle of the impact, and the -extremely
large- phase space available at those energies in terms of
fragmentation channels. We can't possibly
underestimate these aspects of the dynamics,
which are at variance with previous
analysis, where the search for supersymmetric effects
(in the vacuum) seemed
to ignore the fact that our detectors are on the ground and not in space.

For this reason we have resorted to a simple and realistic analysis of
the structure of
airshowers as can be obtained from the current Monte Carlo.

The simplest way to test whether a new interaction at the first proton-proton
impact can have any effect on the shower is to modify the cross section
at the first atmospheric impact using CORSIKA in combination with
some
of the current hadronization models which are supposed to work at and around
the GZK cutoff. There are obvious limitations in this approach, since none of
the
existing codes incorporates any new physics beyond the standard model,
but this is possibly one of the simplest ways to proceed.
For this purpose we have used SYBILL \cite{SIBYLL},
with the appropriate modifications discussed below.

To begin let's start by recalling one feature of the behaviour
of the hadronic
cross section ($p\,p$ or $p\,\bar{p}$) at asymptotically large energies.
There is evidence (see \cite{Ferreiro}) demonstrating
a saturation of the Froissart bound of the total cross section
with rising total energy, $s$. This ${\rm log}^2(s)$ growth
of the total cross section is usually embodied into many of the hadronization
models used in the analysis of first impact and leaves, therefore,
little room for other substantial growth with the opening
of new channels, supersymmetry being one of them.
We should also mention that various significant elaborations \cite{Alan1}
on the growth of the total cross section and the soft pomeron dominance
have been discussed in the last few years and the relation
of this matter \cite{Alan2} with the UHECR events is of utmost relevance.

With this input in mind we can safely
``correct'' the total cross section by at most a (nominal)
factor of 2 and study whether these nominal changes can have any impact on the
structural properties of the showers.

We run simulations on the showers generated by this modification and try
to see whether there is any signal in the multiplicities which points toward a
structural (multiplicity, geometrical) modification of the airshowers in all
or some of its subcomponents. For this purpose we have performed runs at two
different
energies, at the GZK cutoff and 1 decade below, and analyzed the effects
due to these changes.

We show in two figures results on the multiplicities,
obtained at zero zenith angle, of some selected particles
(electrons and positrons, in our case, but similar results hold
for all the dominant components of the final shower) obtained from a large
scale simulation of air showers at and around the
GZK cutoff. We have used the simulation code CORSIKA for this purpose.

\begin{figure}[t]
{\centering
\resizebox*{10cm}{!}{\rotatebox{-90}{\includegraphics{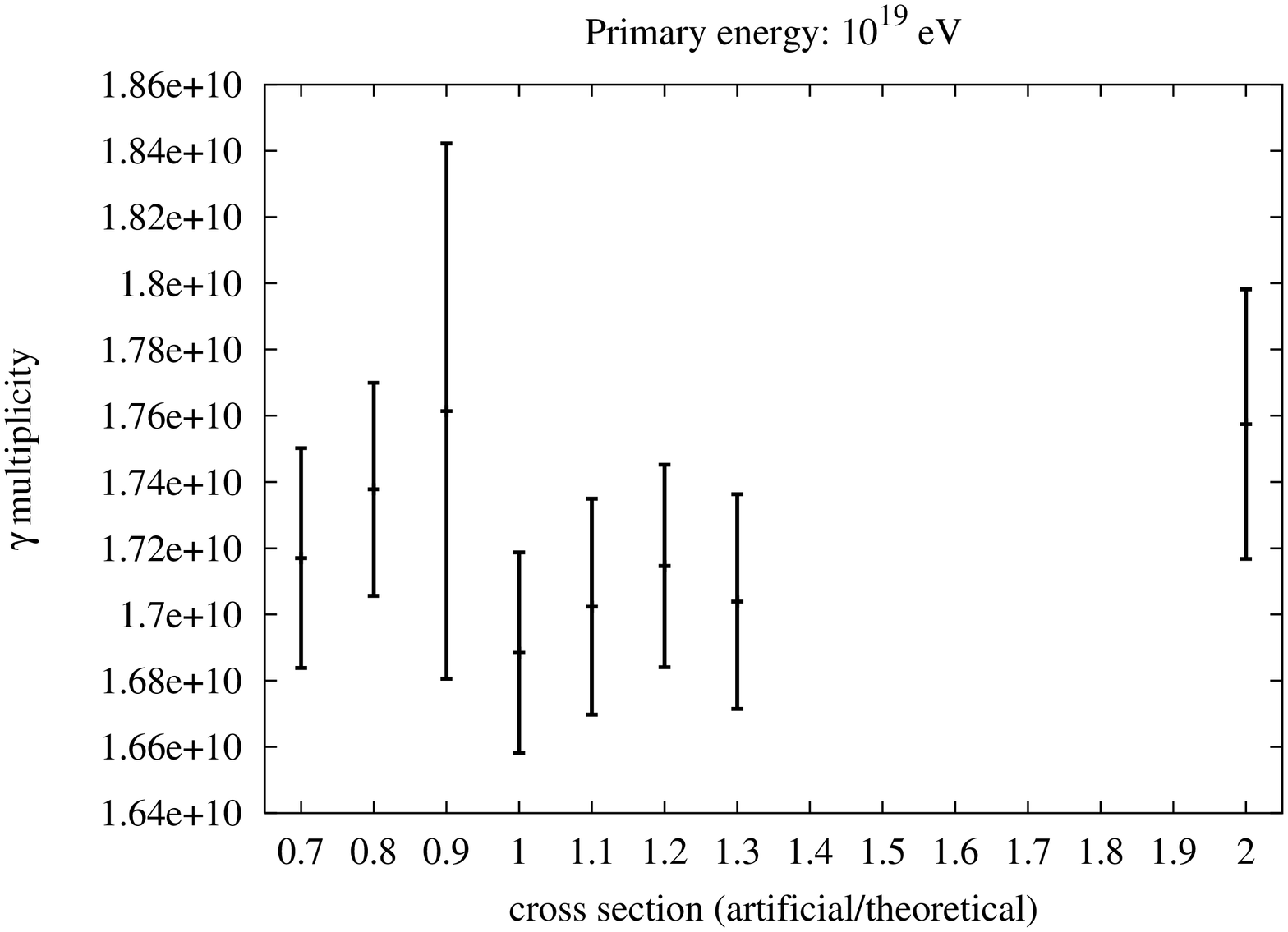}}} \par}
{\centering
\resizebox*{10cm}{!}{\rotatebox{-90}{\includegraphics{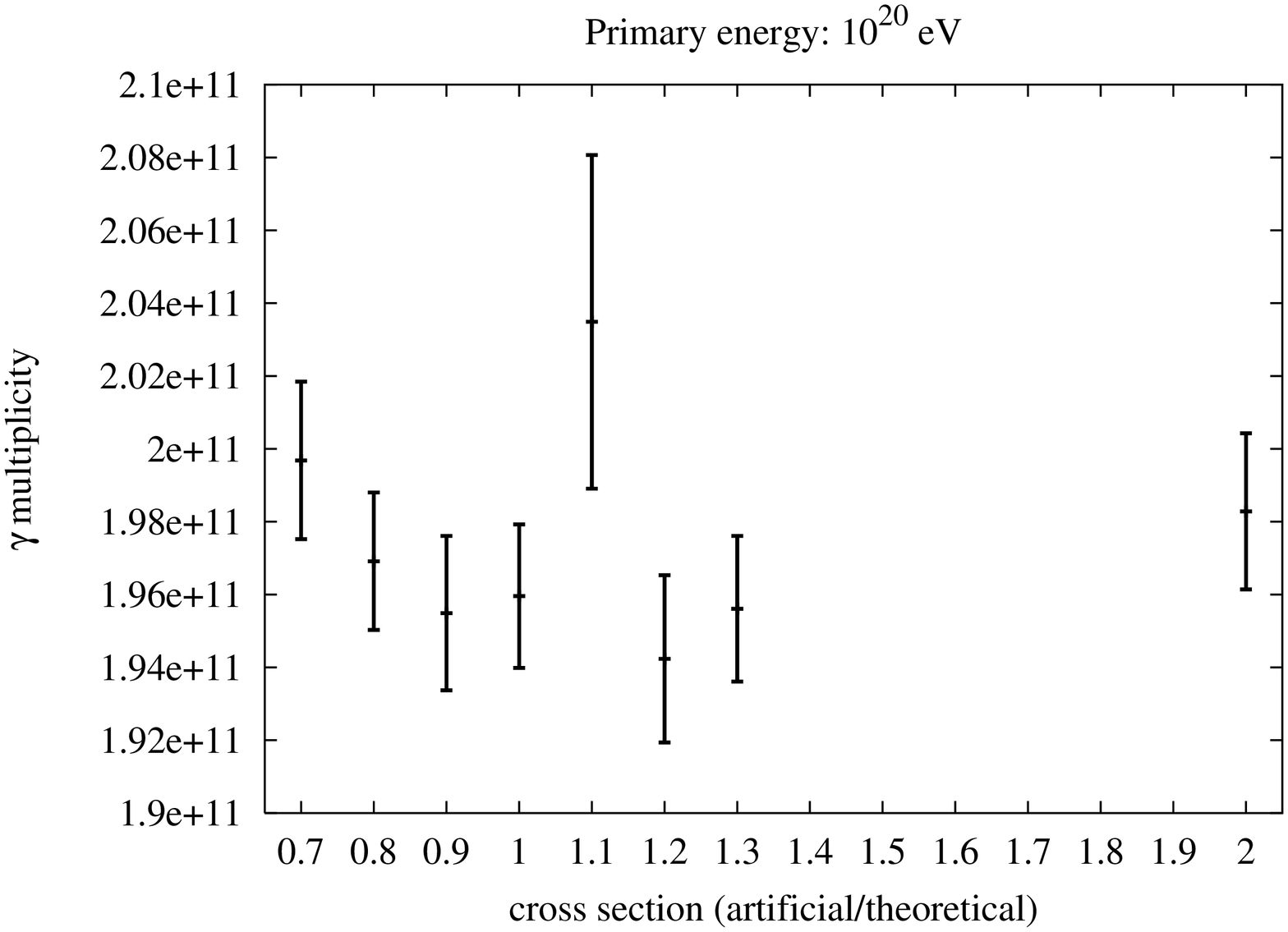}}} \par}
\caption{variation of the photon multiplicity as a function
of the first impact cross section at $10^{19}$ and at $10^{20}$ eV}
\label{multiphoton}
\end{figure}

\begin{figure}[t]
{\centering \resizebox*{10cm}{!}{\rotatebox{-90}{\includegraphics{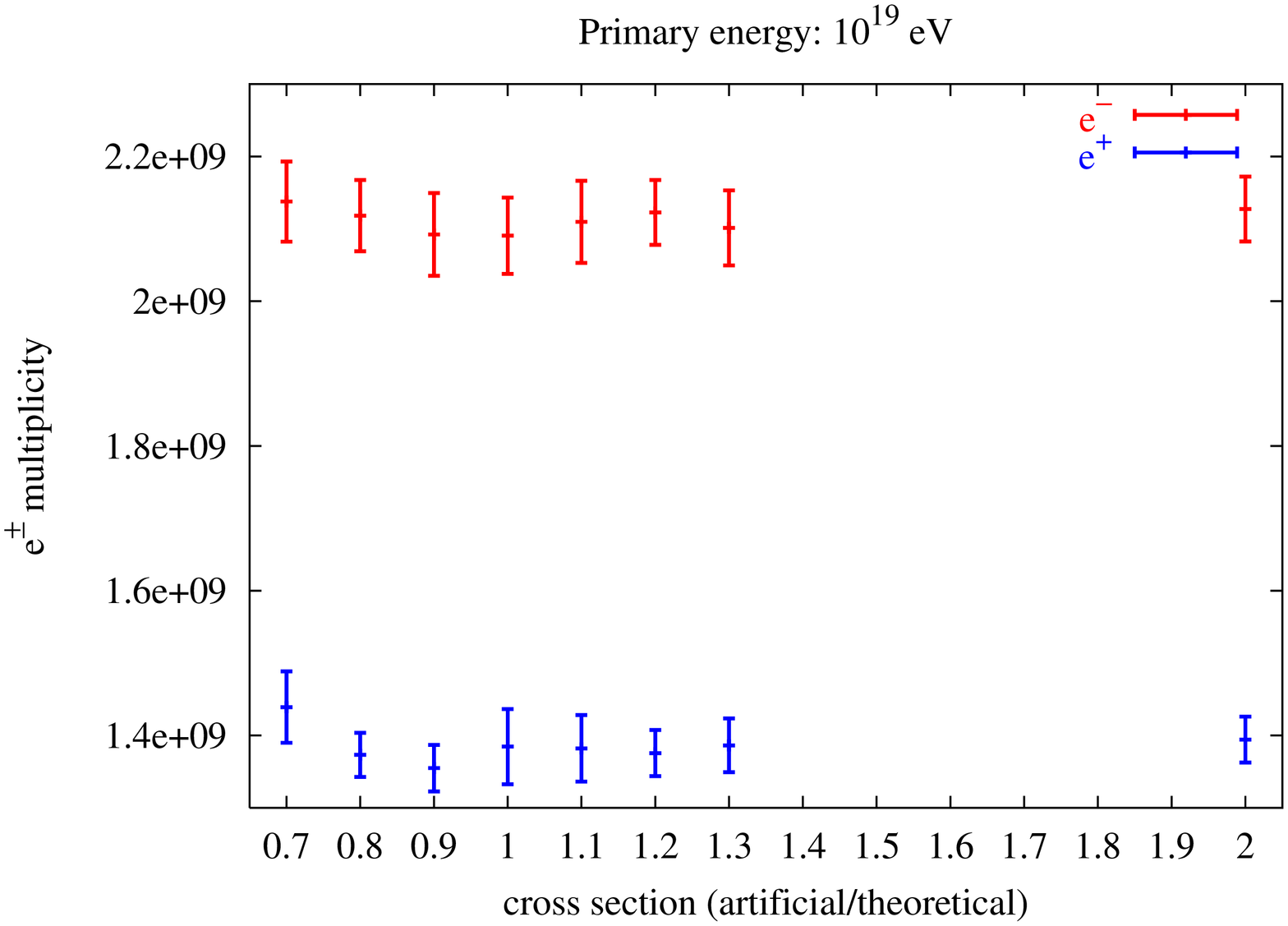}}}
\par}
{\centering \resizebox*{10cm}{!}{\rotatebox{-90}{\includegraphics{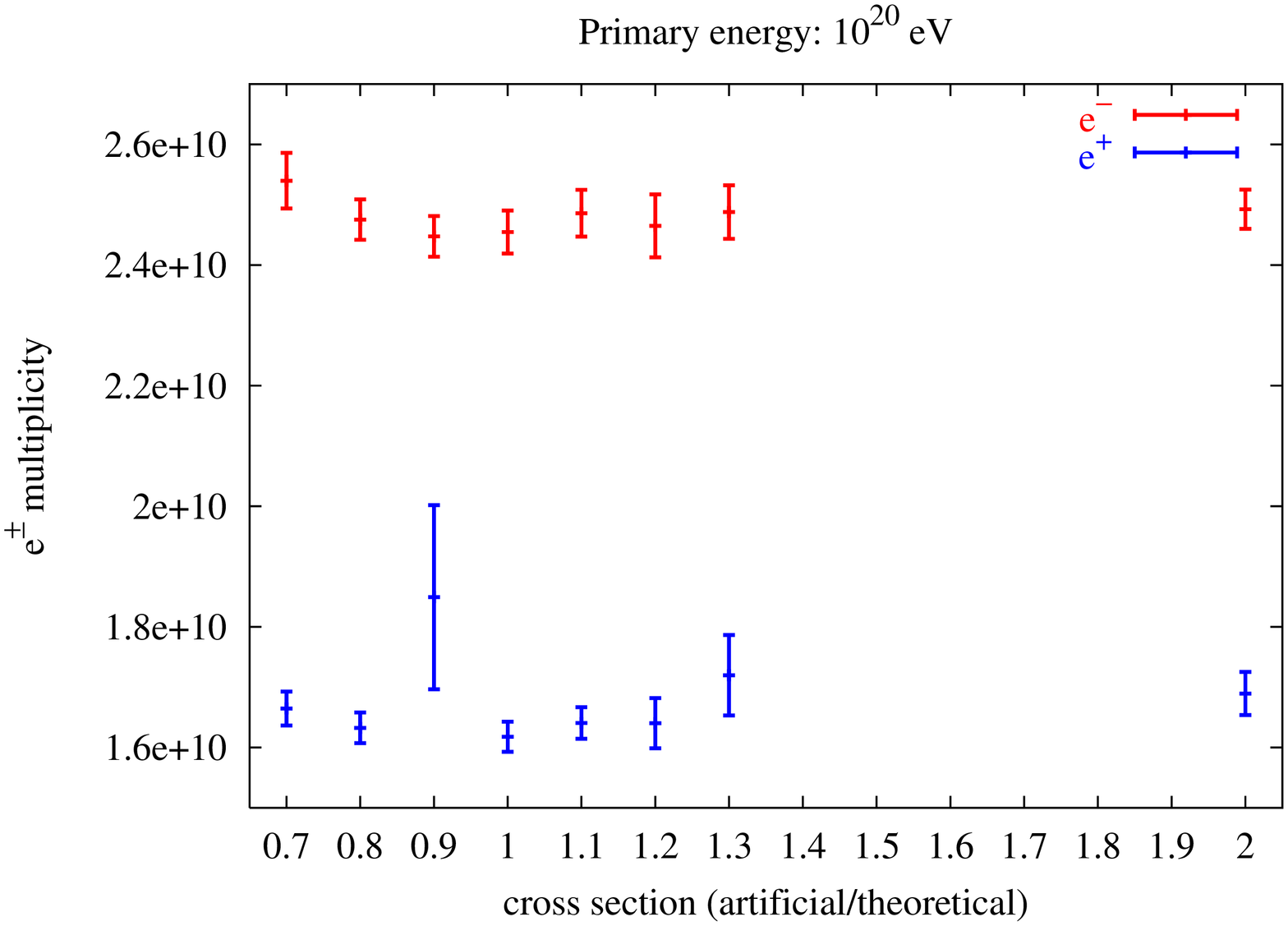}}}
\par}
\caption{variation of the $e^\pm$ multiplicity as a function
of the first impact cross section at $10^{19}$ and at $10^{20}$ eV}
\label{multielectron}
\end{figure}

\begin{figure}[t]
{\centering \resizebox*{10cm}{!}{\rotatebox{-90}{\includegraphics{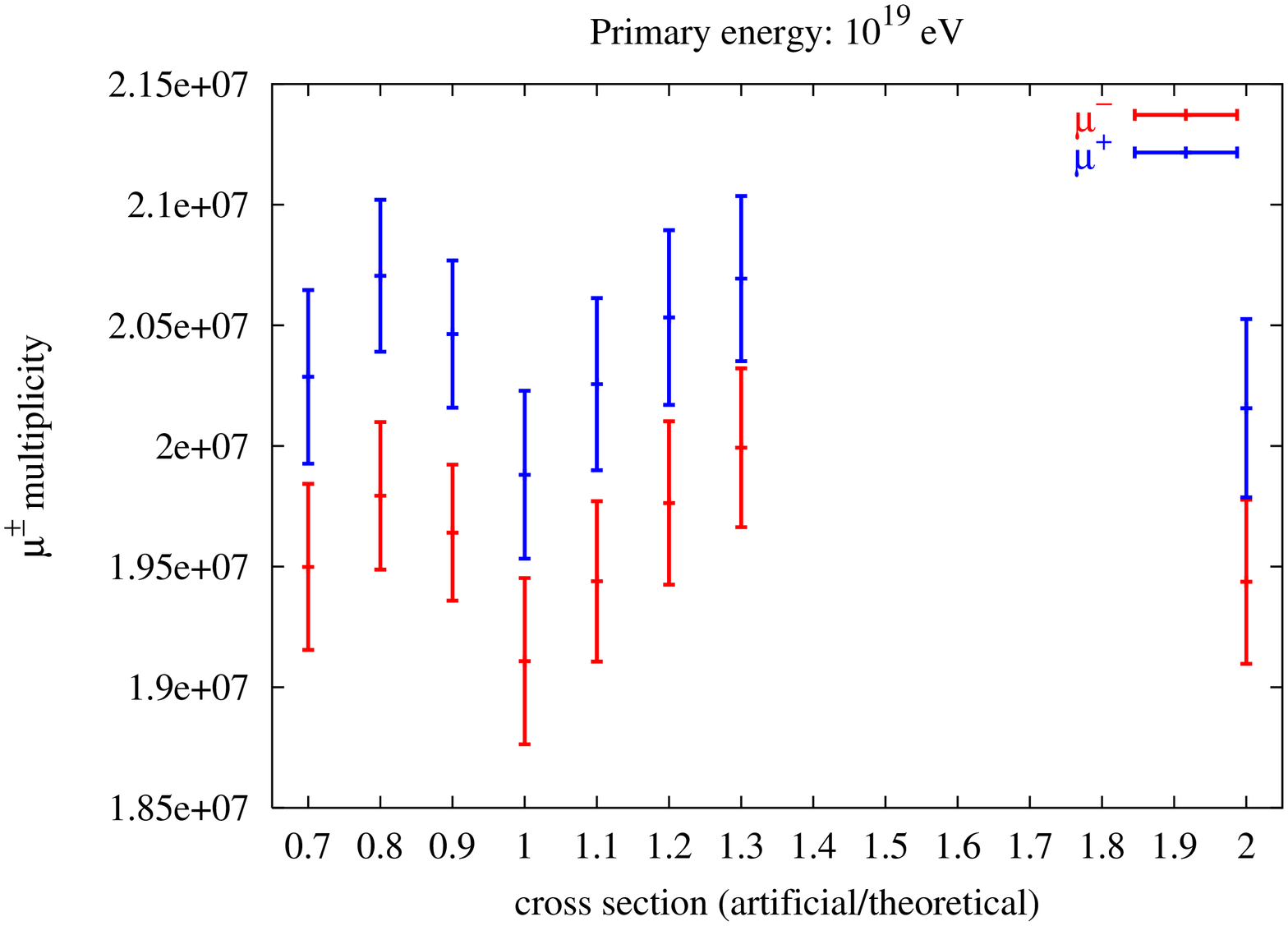}}}
\par}
{\centering \resizebox*{10cm}{!}{\rotatebox{-90}{\includegraphics{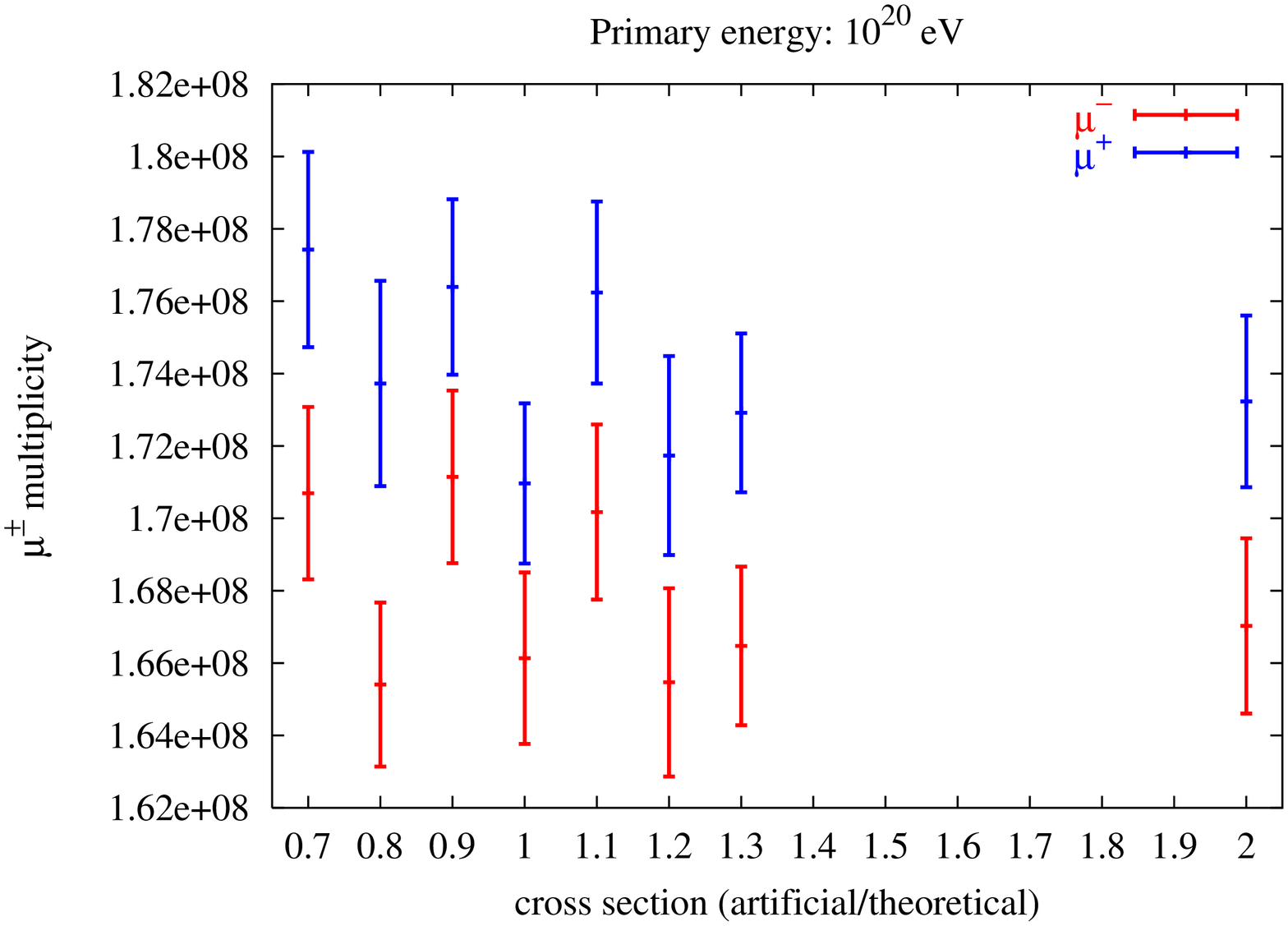}}}
\par}
\caption{variation of the $\mu^\pm$ multiplicity as a function
of the first impact cross section at $10^{19}$ and at $10^{20}$ eV}
\label{multimuon}
\end{figure}

\begin{figure}[t]
{\centering \resizebox*{10cm}{!}{\rotatebox{-90}{\includegraphics{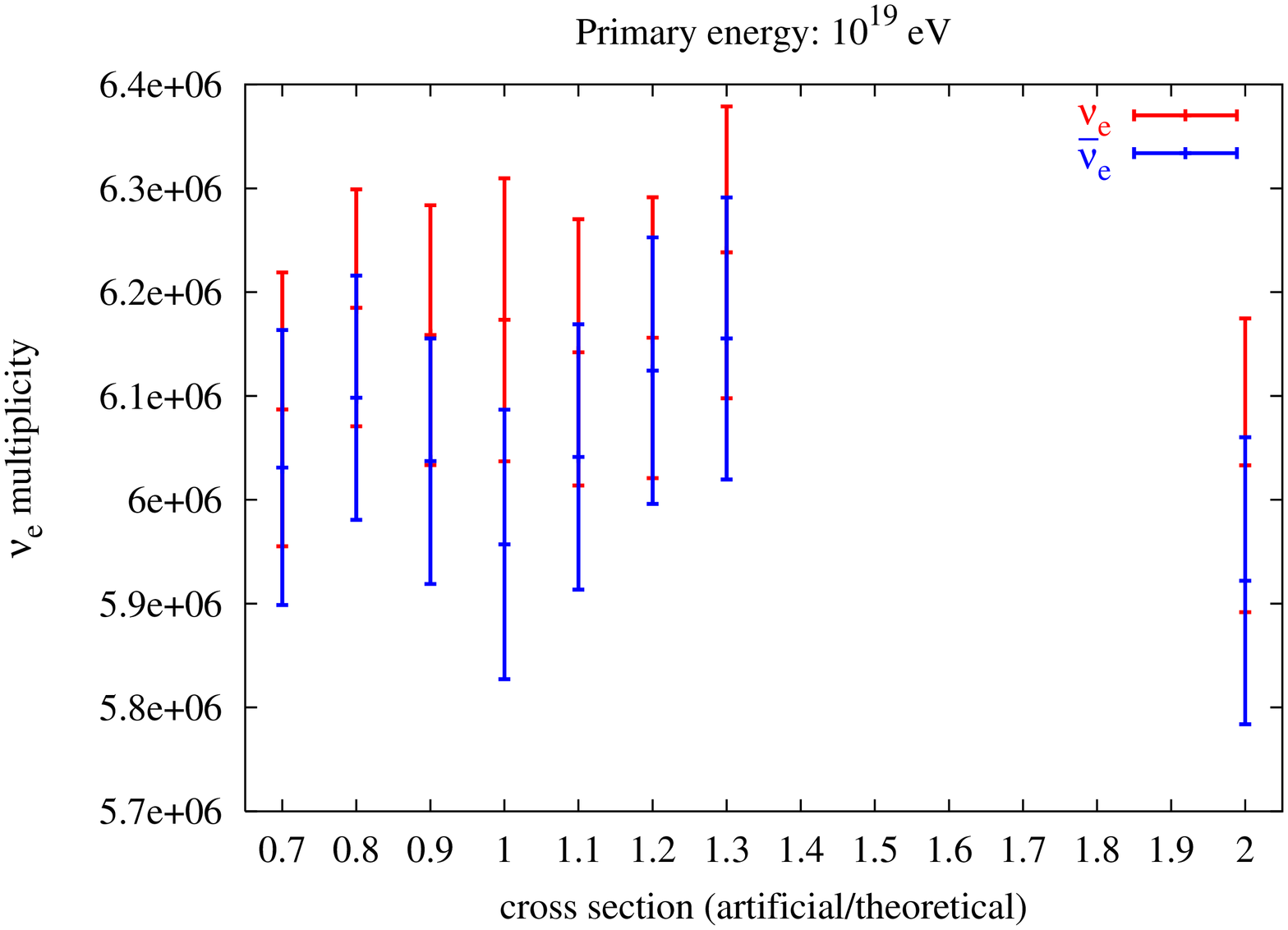}}}
\par}
{\centering \resizebox*{10cm}{!}{\rotatebox{-90}{\includegraphics{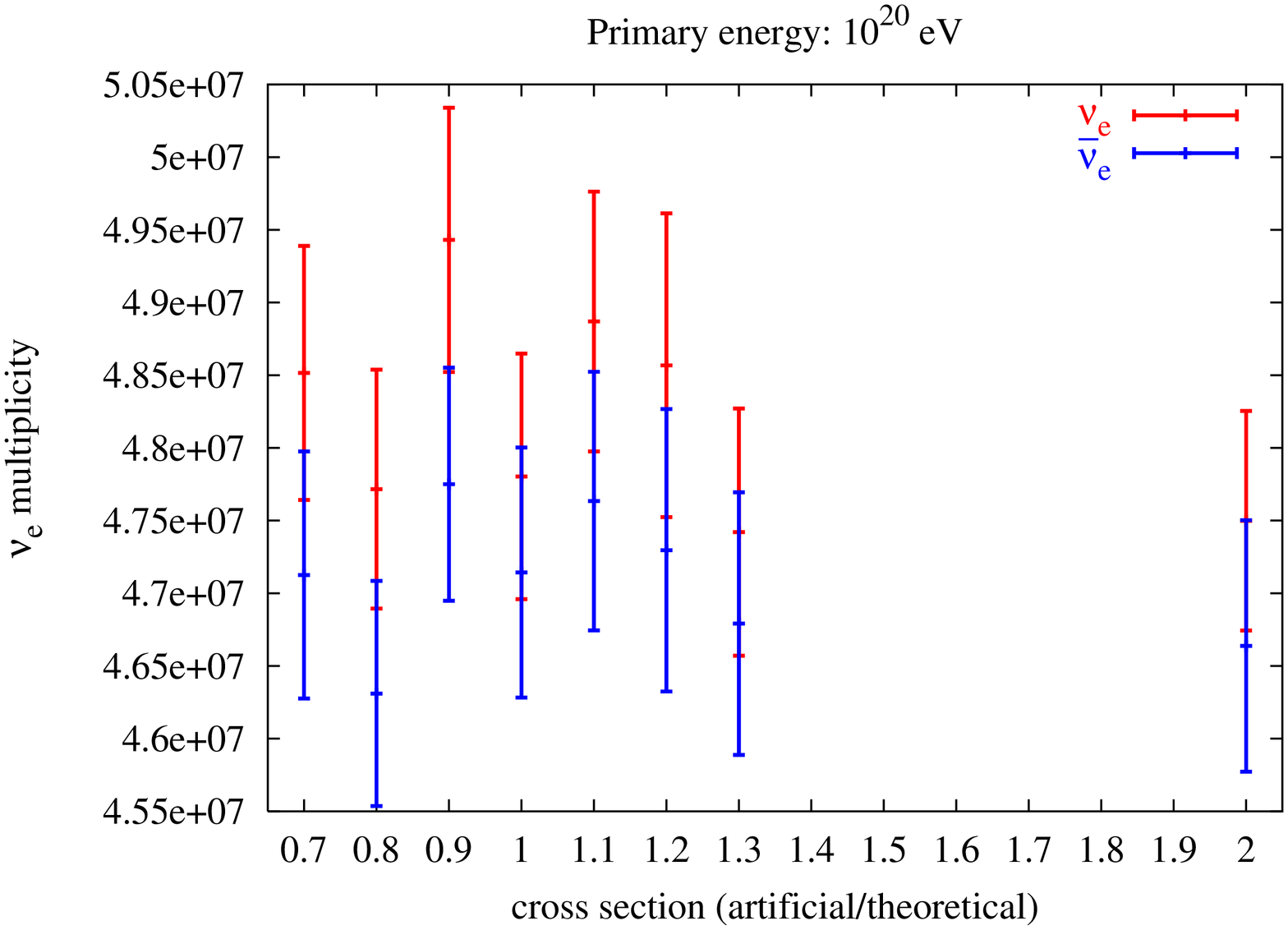}}}
\par}
\caption{variation of the $\nu_e$ multiplicity as a function
of the first impact cross section at $10^{19}$ and at $10^{20}$ eV}
\label{multielectronneutrino}
\end{figure}

\begin{figure}[t]
{\centering
\resizebox*{10cm}{!}{\rotatebox{-90}{\includegraphics{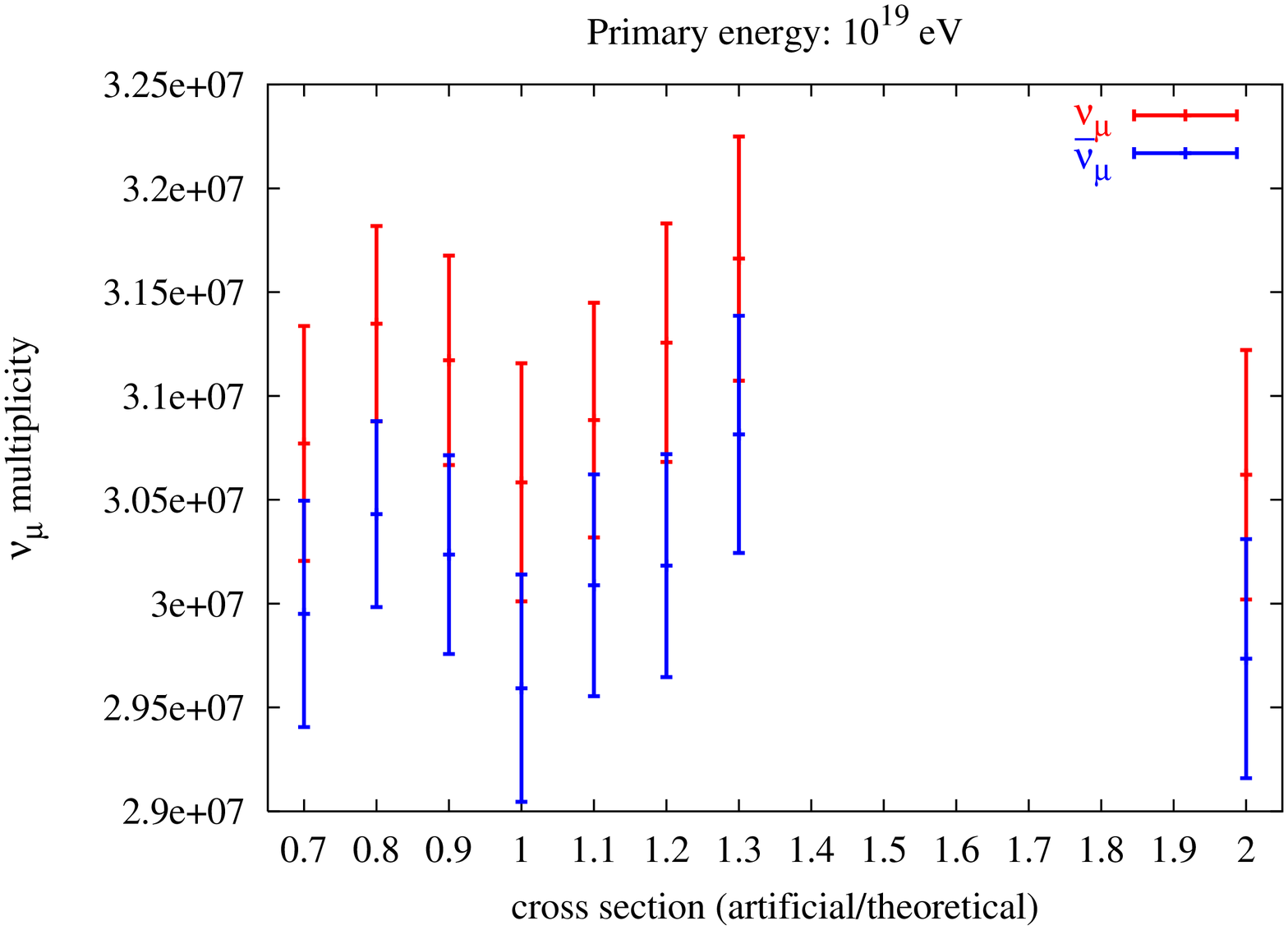}}} \par}
{\centering
\resizebox*{10cm}{!}{\rotatebox{-90}{\includegraphics{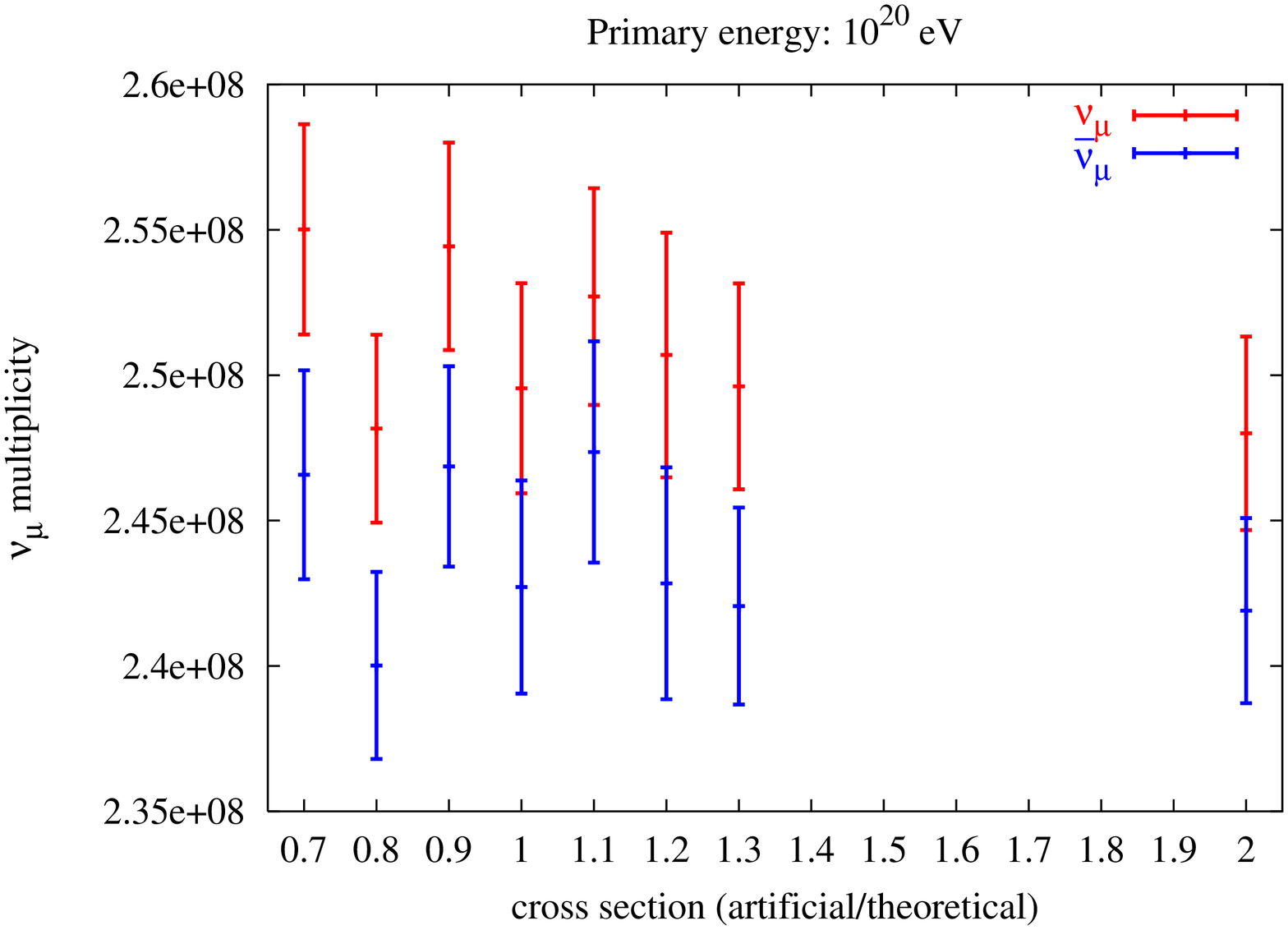}}} \par}
\caption{variation of the $\nu_\mu$ multiplicity as a function
of the first impact cross section at $10^{19}$ and at $10^{20}$ eV}
\label{multimuonneutrino}
\end{figure}

In Figs.~\ref{multiphoton}-\ref{multimuonneutrino}
we show plots obtained simulating an artificial first proton
impact in which we have modified the first interaction cross section by a
nominal factor ranging from 0.7 to 2. We plot on the y-axis the corresponding
fluctuations in the multiplicities both for electrons and positrons.
Statistical fluctuations \footnote{we keep the height of the first proton
impact with the atmosphere arbitrary for each selected correction factor
(x-axis)}
have been estimated using bins of 80 runs. The so-developed showers have been
thinned using the Hillas
algorithm, as usually done in order to make the results of these simulations
manageable,
given the size of the showers at those energies. As one can immediately see,
the artificial
corrections on the cross section are compatible with ordinary fluctuations of
the air-shower. We have analyzed all the major subcomponents of the air shower,
photons and leptons,
together with the corresponding neutrino components.
We can summarize these findings by assessing that a modified first impact,
at least for such correction factors in the 0.7-2 range
in the cross section, are unlikely to modify the multiplicities in any
appreciable way.

\begin{figure}[t]
{\centering
\resizebox*{10cm}{!}{\rotatebox{-90}{\includegraphics{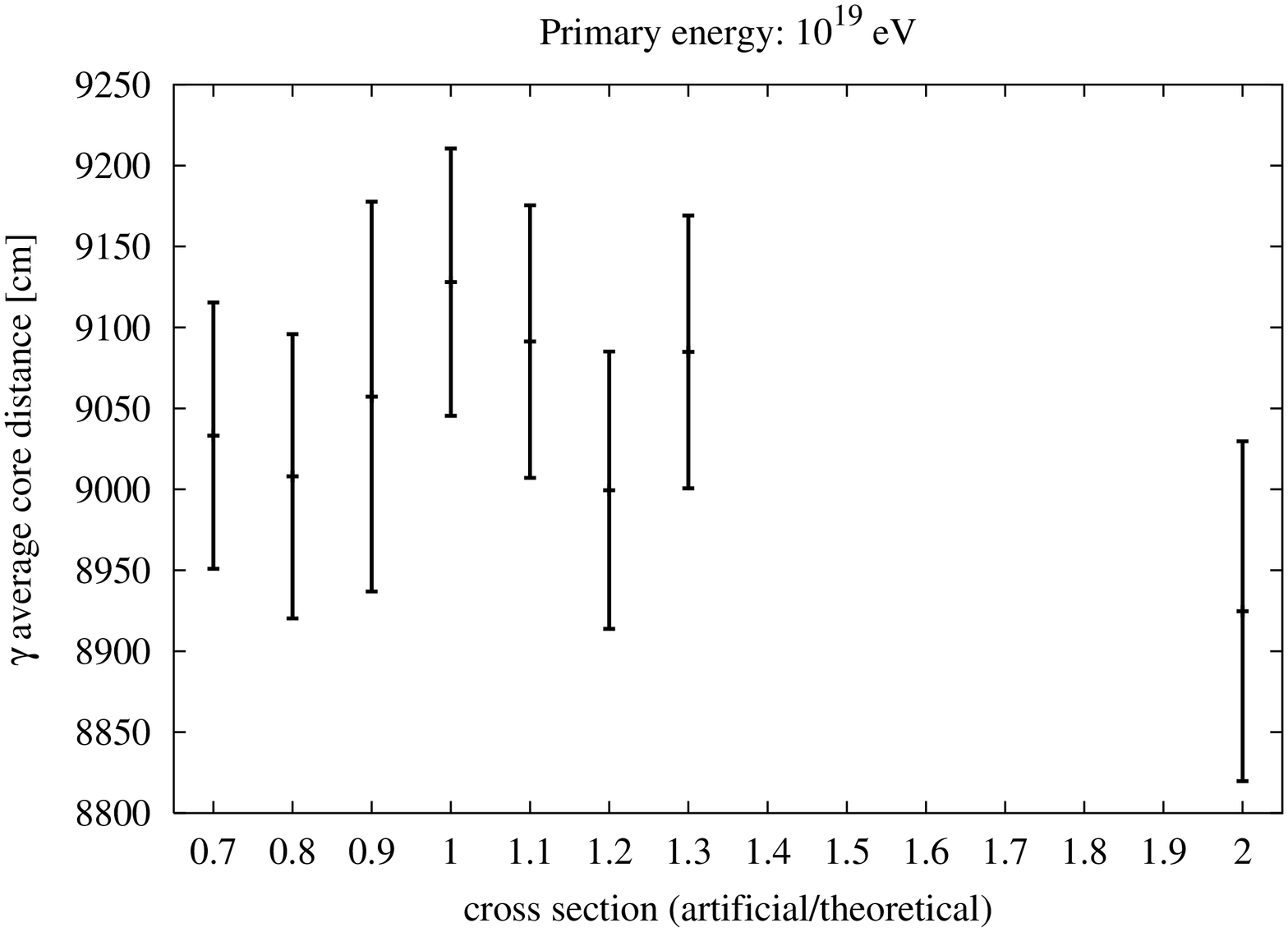}}} \par}
{\centering
\resizebox*{10cm}{!}{\rotatebox{-90}{\includegraphics{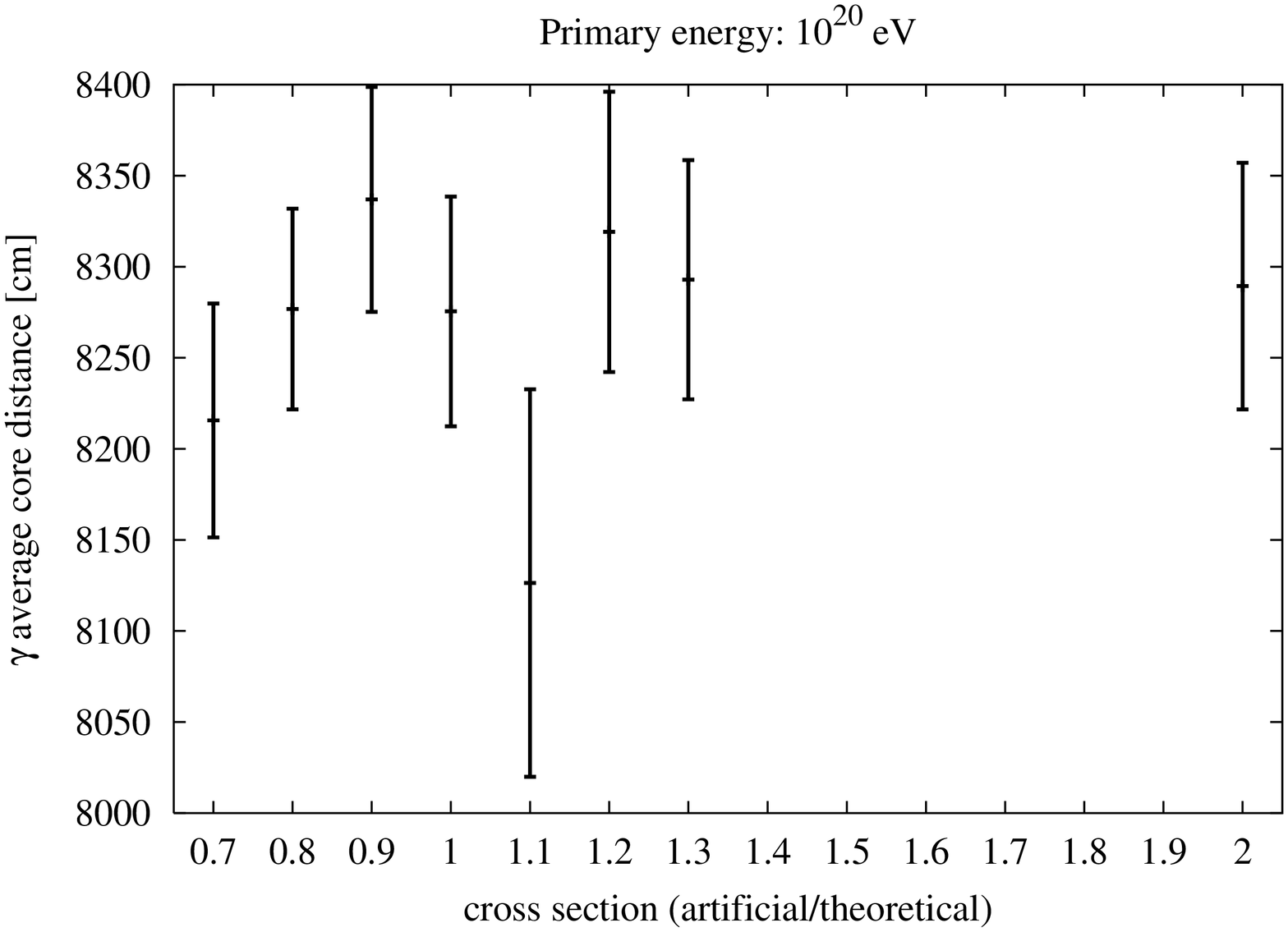}}} \par}
\caption{Lateral distributions of photons as a function of the first impact
cross section at $10^{19}$ and at $10^{20}$ eV}
\label{lateralphoton}
\end{figure}

\begin{figure}[t]
{\centering \resizebox*{10cm}{!}{\rotatebox{-90}{\includegraphics{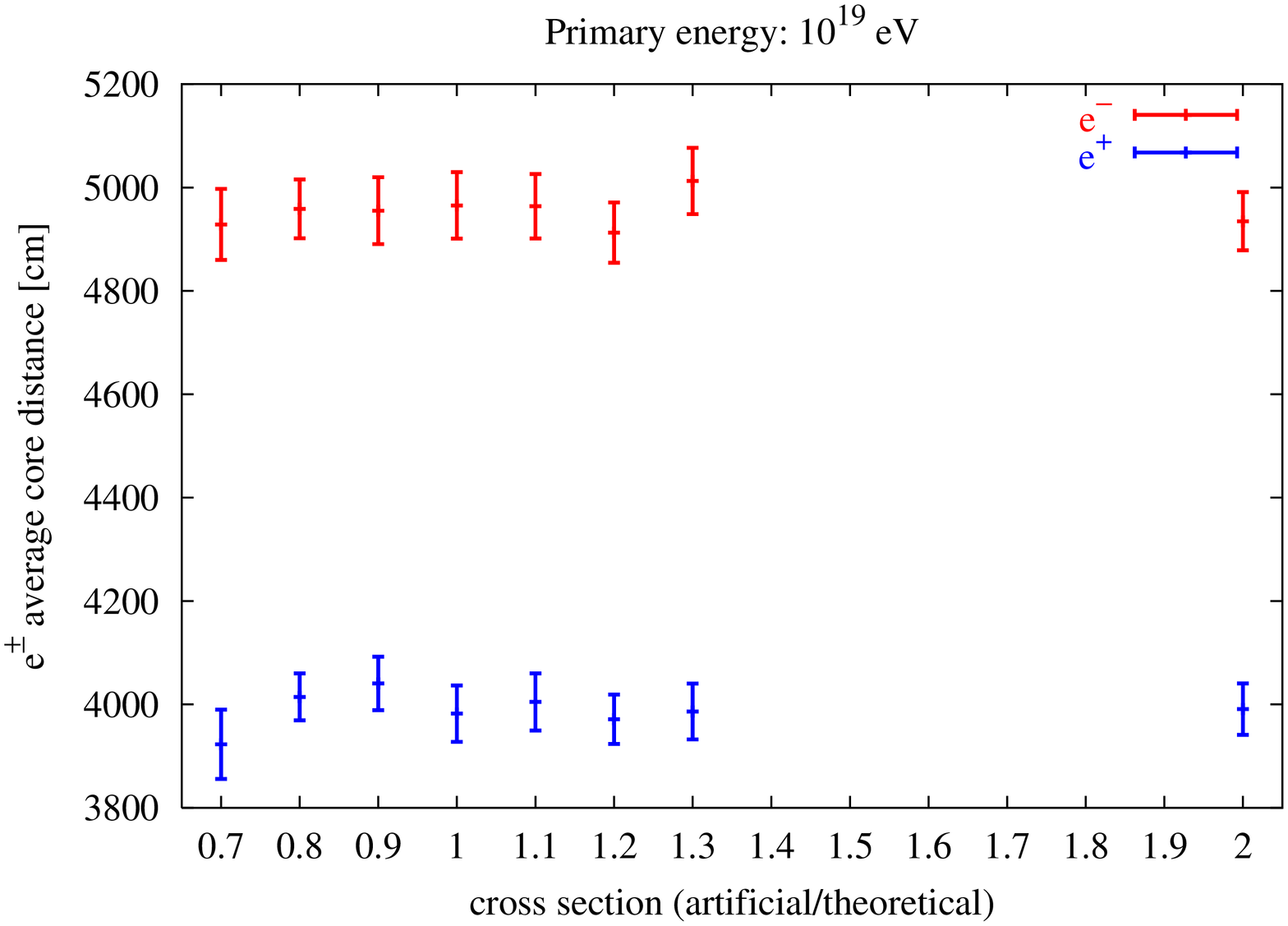}}}
\par}
{\centering \resizebox*{10cm}{!}{\rotatebox{-90}{\includegraphics{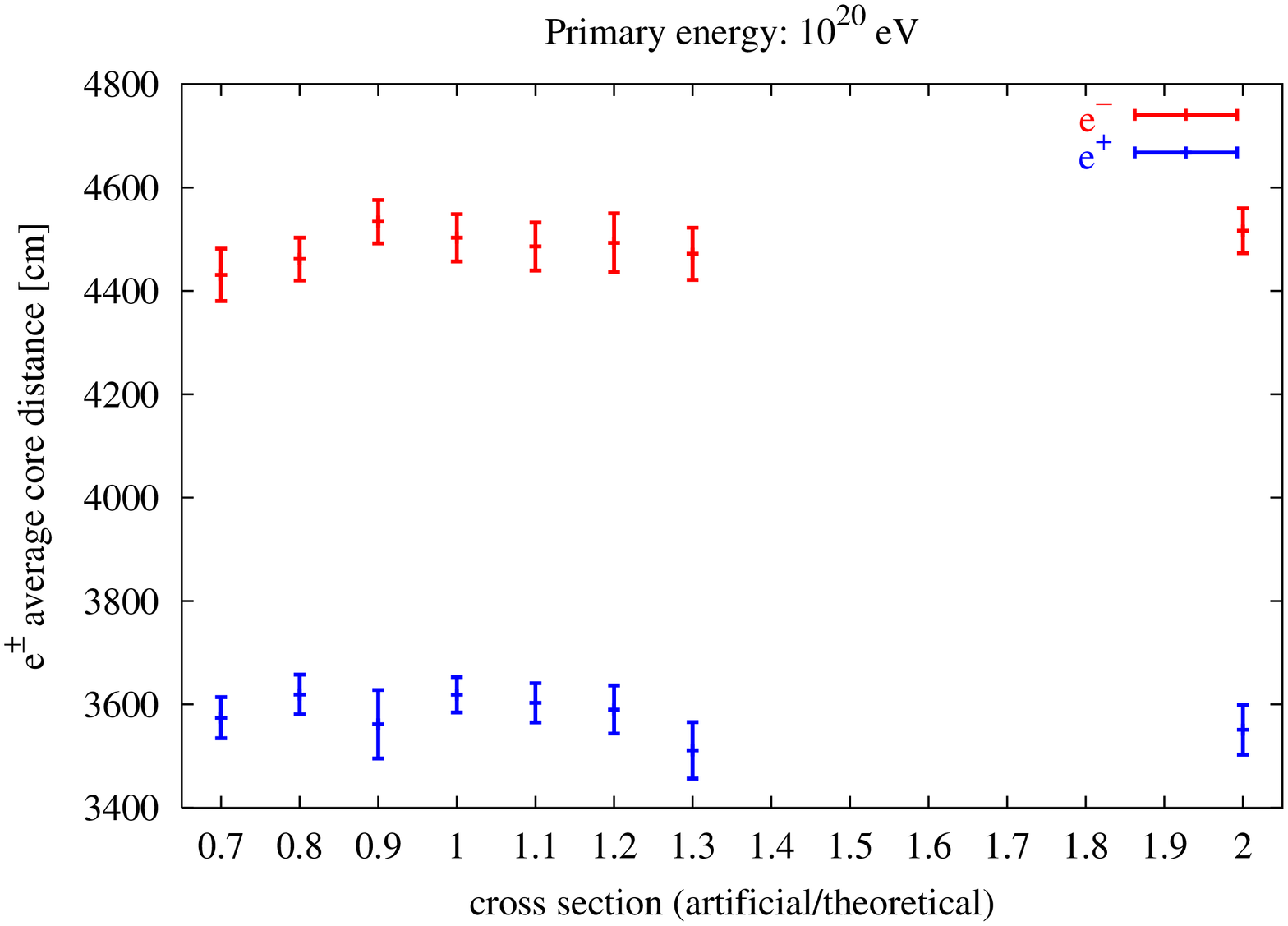}}}
\par}
\caption{Lateral distributions of $e^\pm$
as a function of the first impact cross section at $10^{19}$ and at $10^{20}$
eV}
\label{lateralepm}
\end{figure}

\begin{figure}[t]
{\centering \resizebox*{10cm}{!}{\rotatebox{-90}{\includegraphics{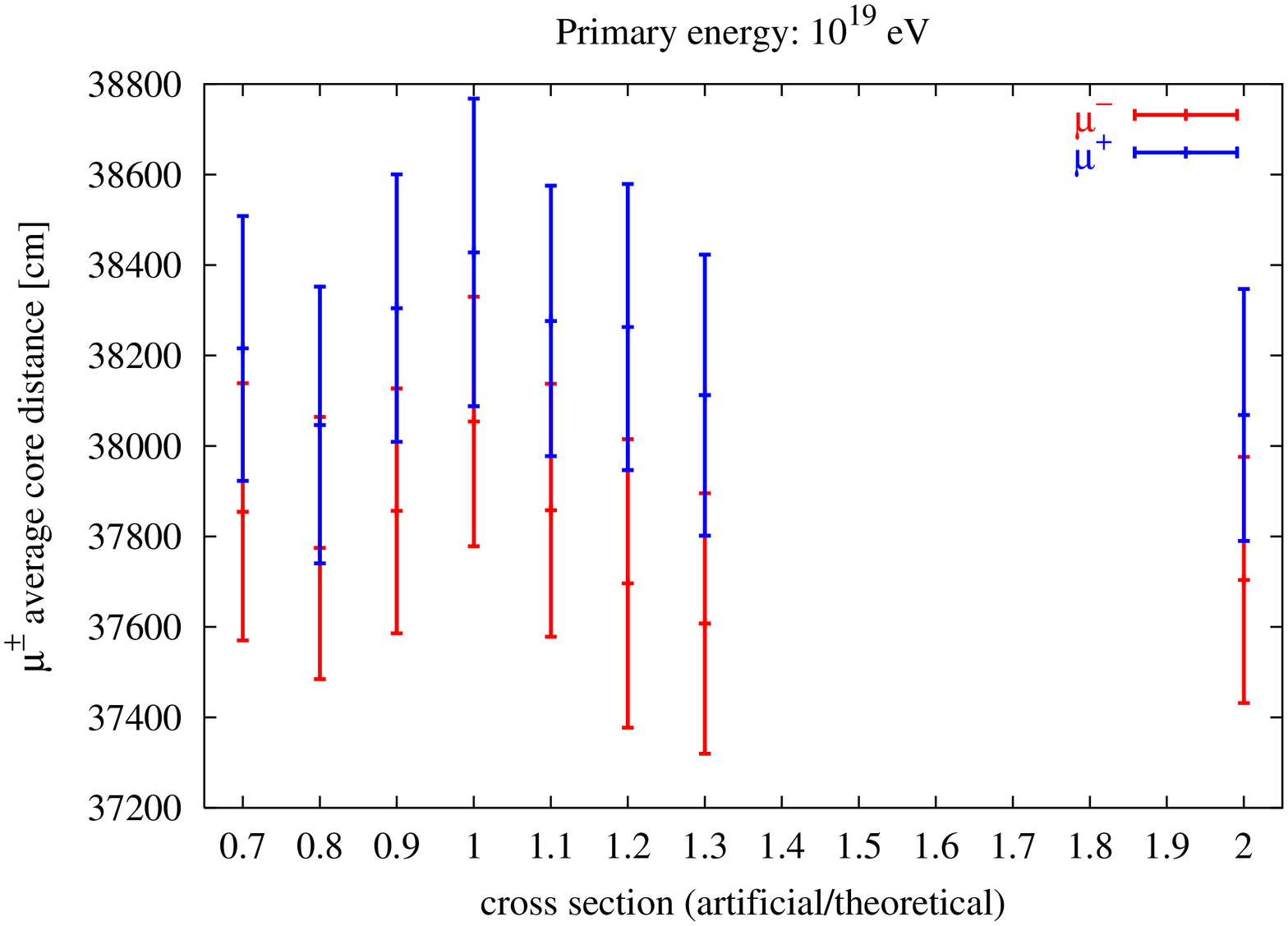}}}
\par}
{\centering \resizebox*{10cm}{!}{\rotatebox{-90}{\includegraphics{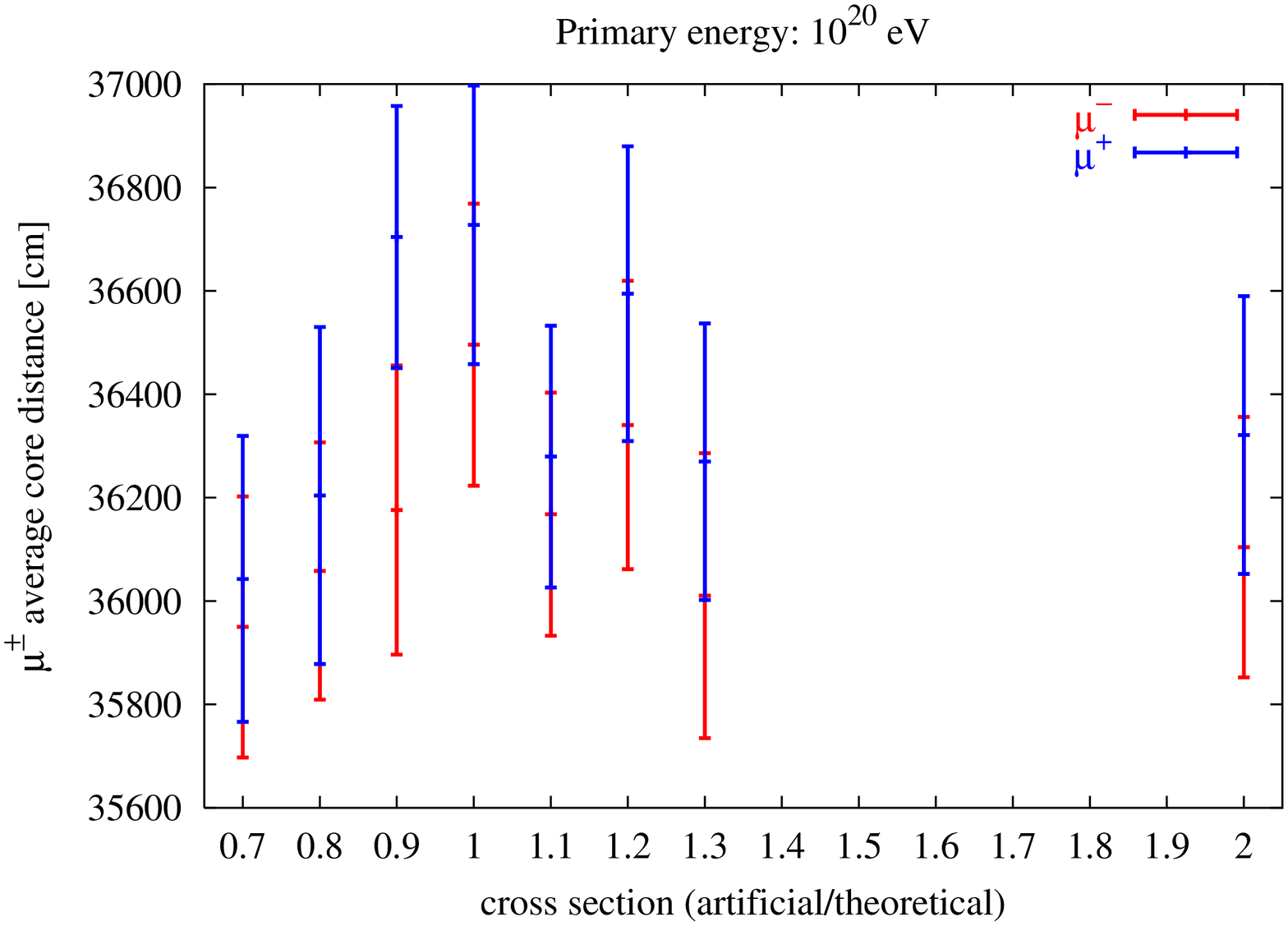}}}
\par}
\caption{Lateral distributions of $\mu^\pm$
as a function of the first impact cross section at $10^{19}$ and at $10^{20}$
eV}
\label{lateralmupm}
\end{figure}

\begin{figure}[t]
{\centering \resizebox*{10cm}{!}{\rotatebox{-90}{\includegraphics{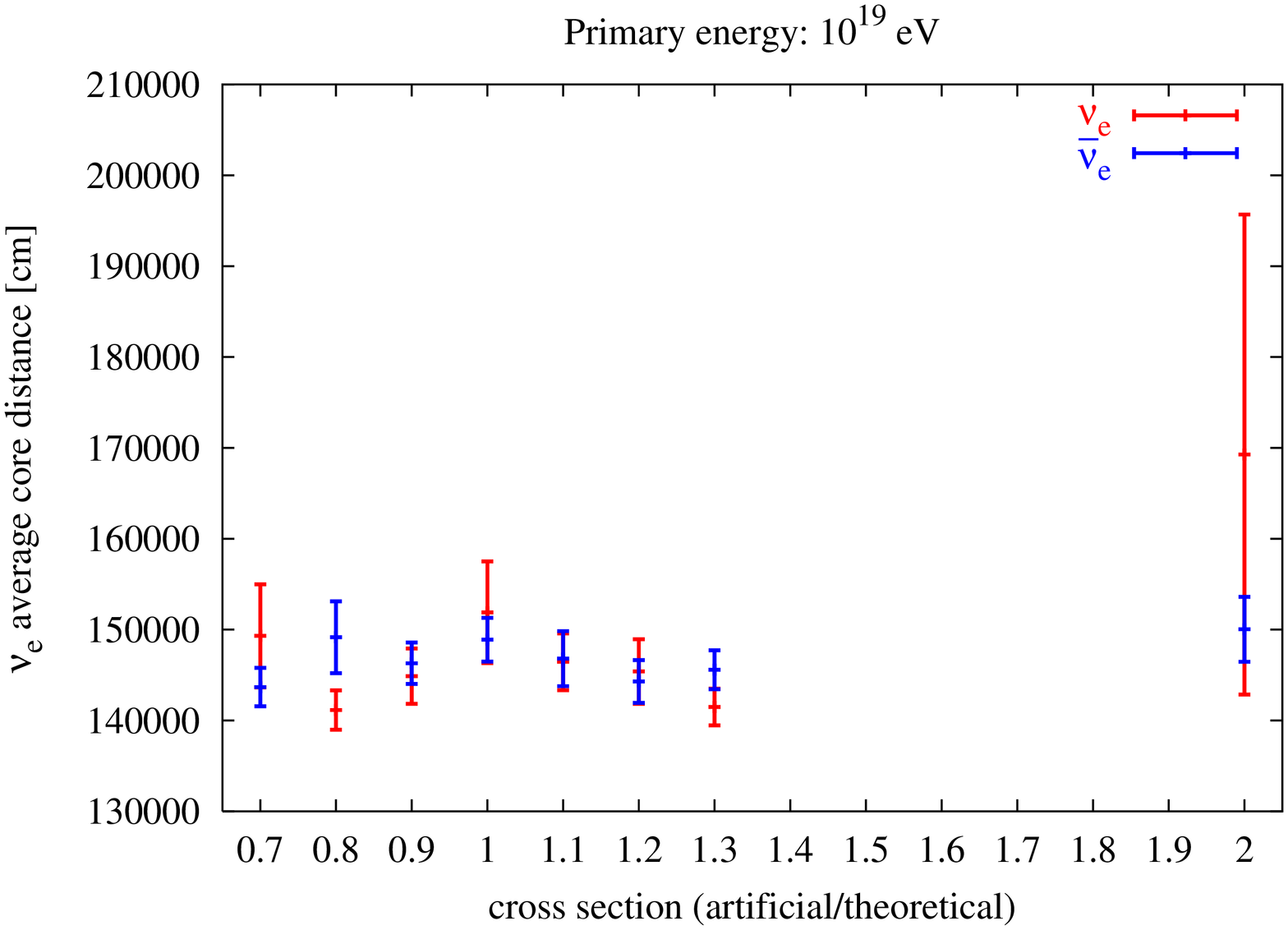}}}
\par}
{\centering \resizebox*{10cm}{!}{\rotatebox{-90}{\includegraphics{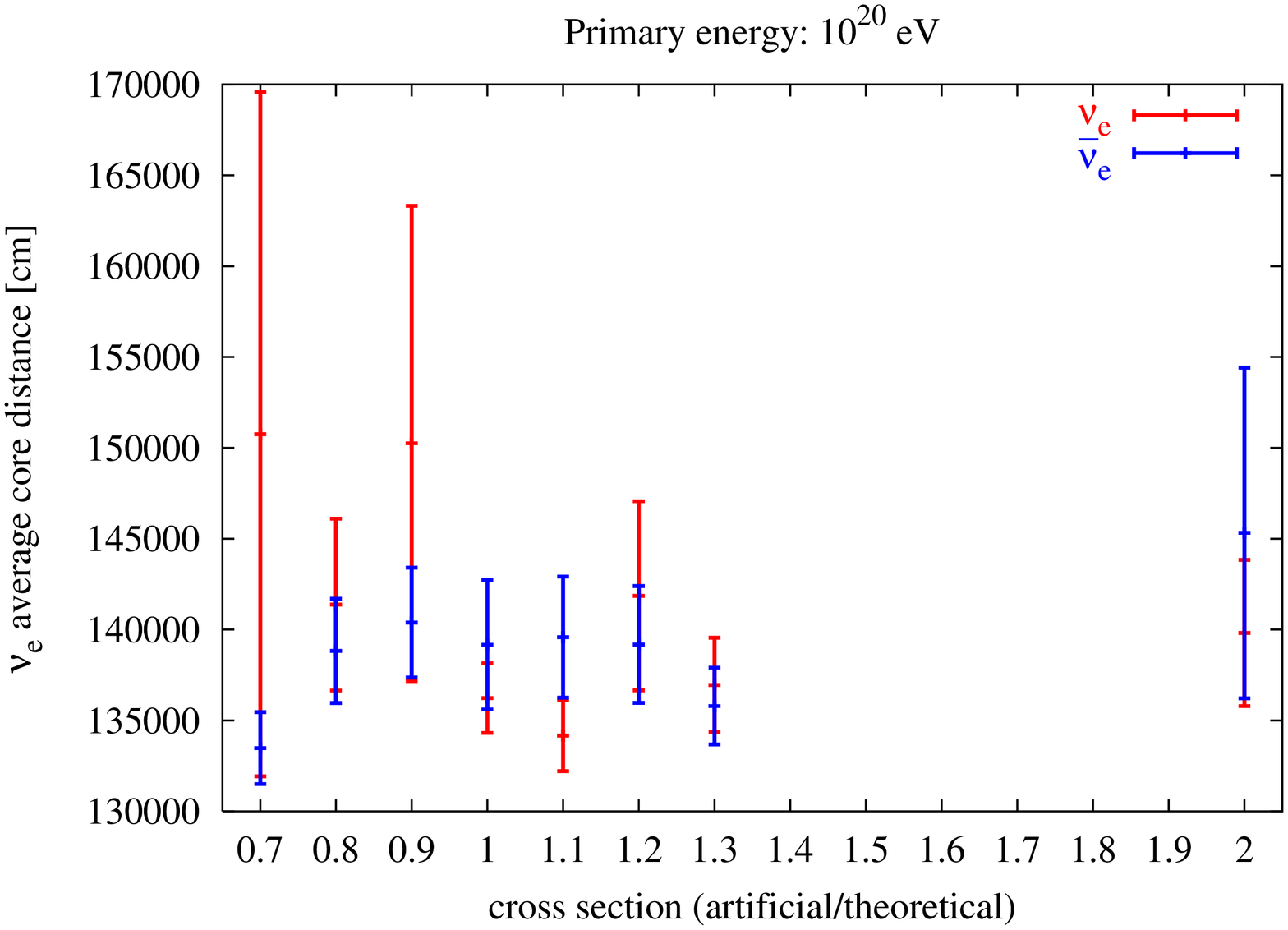}}}
\par}
\caption{Lateral distributions of $\nu_e$
as a function of the first impact cross section at $10^{19}$ and at $10^{20}$
eV}
\label{lateralnue}
\end{figure}

\begin{figure}[t]
{\centering
\resizebox*{10cm}{!}{\rotatebox{-90}{\includegraphics{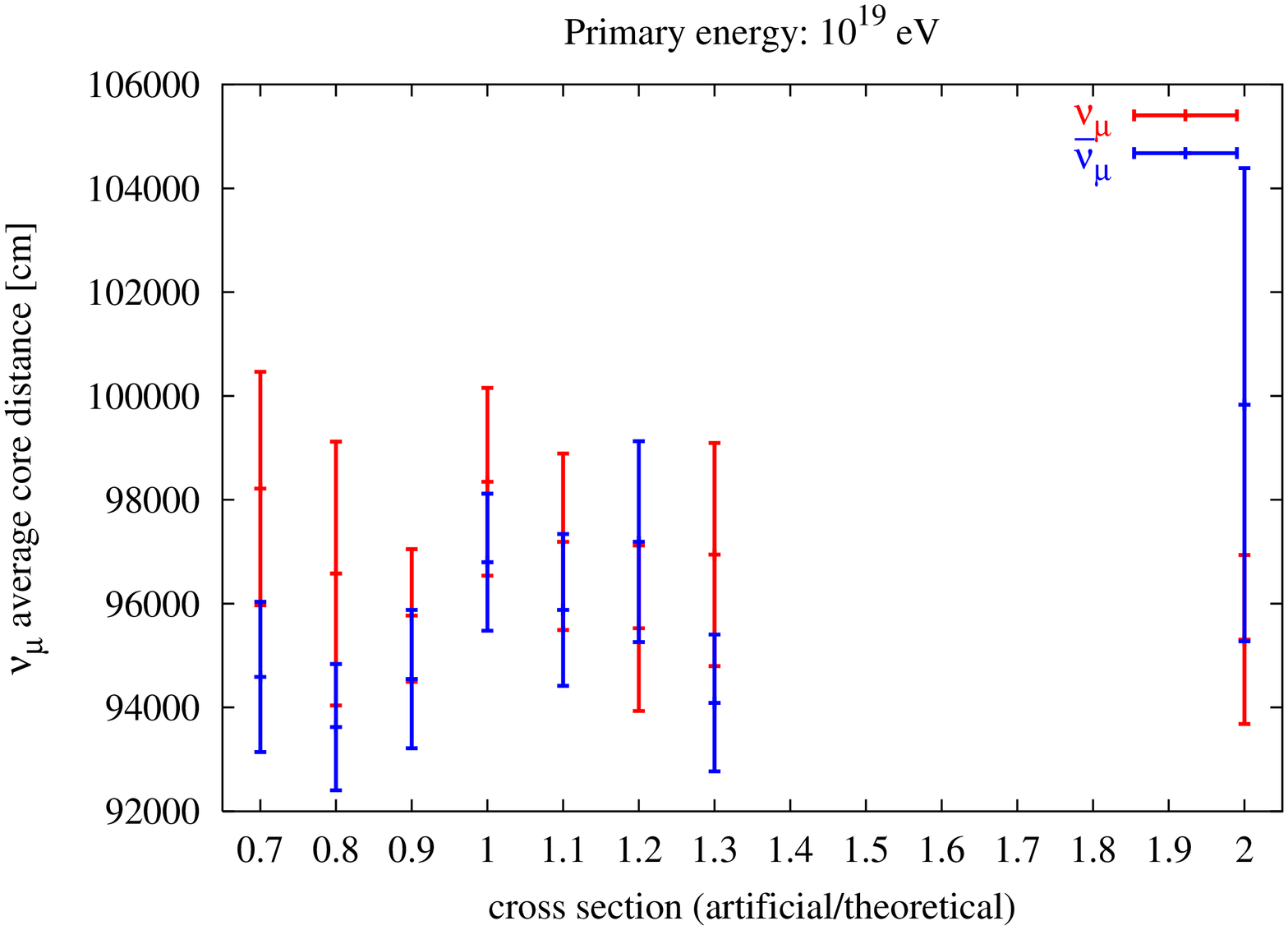}}} \par}
{\centering
\resizebox*{10cm}{!}{\rotatebox{-90}{\includegraphics{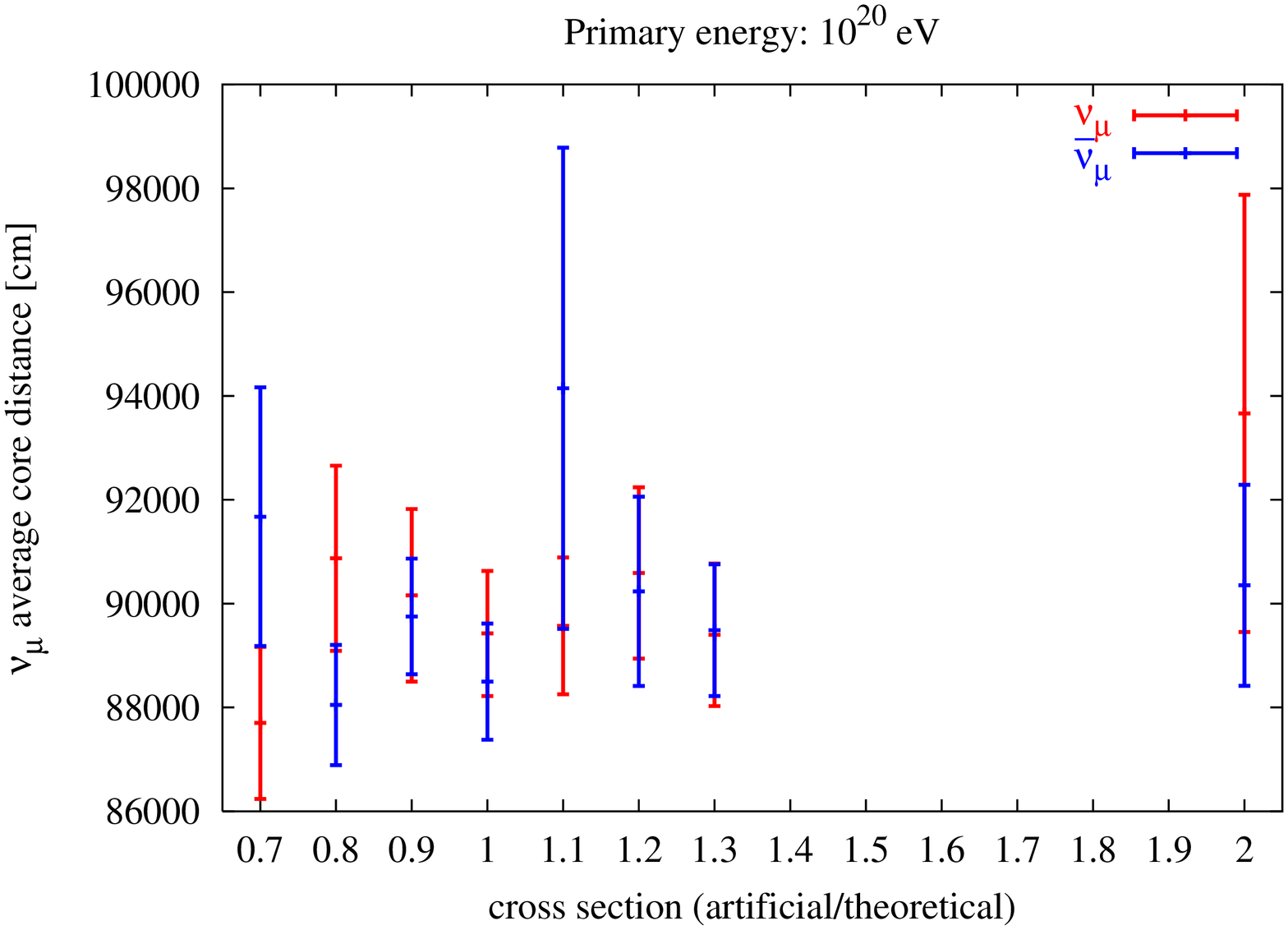}}} \par}
\caption{Lateral distributions of $\nu_\mu$
as a function of the first impact cross section at $10^{19}$ and at $10^{20}$
eV}
\label{lateralnumu}
\end{figure}

A second test is illustrated in Figs.~\ref{lateralphoton}-\ref{lateralnumu}.
Here we plot the same correction factors on the
x-axis as in the previous plots (\ref{multiphoton}-\ref{multimuonneutrino})
but we show on the y-axis (for the same particles) the average point
of impact on the detector and its corresponding statistical fluctuations.
As we increase the correction
factor statistical fluctuations in the formation of the air shower
seem to be compatible with the modifications induced by the ``new physics''
of the first impact and no special new effect is observed.

Fluctuations of these type, generated by a minimal modification of the existing
codes
only at the first impact may look simplistic, and can possibly
be equivalent to ordinary simulations with a simple rescaling
of the atmospheric height at which the first collision occurs, since the
remaining
interactions are, in our approach, unmodified.
The effects we have been looking for, therefore, appear subleading compared to
other standard fluctuations which take place in the formation of the cascade.
On the other hand, drastic changes in the
structure of the air shower should possibly depend mostly on the physics
of the first impact and only in a less relevant way on the modifications
affecting the
cascade that follows up.
We have chosen to work at an energy of $10^{20}$ eV but we do not
observe any substantial modifications of our results at lower energies
($10^{19}$ eV), except for the multiplicities which are down by a factor
of 10. Our brief analysis, though simple, has the purpose to
illustrate one of the many issues which we believe
should be analyzed with great care in the near future: the physics of the first
impact and substantial additional modifications to the existing codes
in order to see whether any new physics can be extracted from
these measurements.

\section{Conclusions}
We have tried to analyze with a searching criticism the possibility of
detecting new physics at AUGER using current ideas about supersymmetry,
the QCD evolution and such.
While the physics possibilities of these experiments
are far reaching and may point toward a validation
or refusal of the existence of a GZK cutoff, we have argued that
considerable progress still needs to be done
in order to understand better the hadronization models at very large
energy scales. Our rather conservative viewpoint stems from the fact
that the knowledge of the
structure of the hadronic showers at large energies is still
under debate and cannot be conclusive.
We have illustrated by an extended and updated simulation some of the
characteristics of the showers, the intrinsic fluctuations in the lateral
distributions, the multiplicities of the various subcomponents and of the
total spectrum, under some realistic conditions. We have also tried
to see whether nominal and realistic changes in the cross section of the
first impact may affect the multiplicities, with a negative outcome.
We have however pointed out, in a positive way, that new physics models
predicting large missing
energy may have a chance to be identified, since the trend followed by the
total multiplicities
(in a log-log scale) appear to be strikingly linear.

We do believe, however, that other and even more extensive simulation studies
should be done, in combination with our improved understanding of small-x
effects
in QCD at large parton densities, in order to further enhance the physics
capabilities
of these experiments. Another improvement in the extraction of new physics
signals from the
UHECR experiments will come from the incorporation of new physics in the
hadronization
models.

\centerline{\bf Acknowledgments}
We thank D. Martello and Philippe Jetzer for discussions and Prof. D. Heck
for correspondence.
The work of A.C. and C.C. is supported in part by INFN
(iniziativa specifica BA-21). The work of A.F. is supported in part by PPARC.
Simulations have been performed using the INFN-LECCE/Astrophysics
computational cluster. C.C. thanks the University of Oxford and
the Theory Group at the University. of Zurich (Irchel)
for hospitality while completing this work.


\clearpage

\newpage

\end{document}